\documentclass[12pt,preprint]{aastex}
\usepackage{pjournal}
\shorttitle{Predictions for Cosmological Infrared Surveys from Space
with MIPS/SIRTF}
\shortauthors{H. Dole et al.}

\begin{document}

\title{Predictions for Cosmological Infrared Surveys from Space with
the Multiband Imaging Photometer for SIRTF (MIPS).} 

\author{H. Dole}
\affil{Steward Observatory, University of Arizona, 933 N Cherry Ave,
Tucson, AZ 85721, USA}
\email{hdole@as.arizona.edu}

\author{G. Lagache and J.-L. Puget}
\affil{Institut d'Astrophysique Spatiale, b\^at 121, Universit\'e
Paris Sud, F-91405 Orsay Cedex, France}

\begin{abstract}
We make predictions for the cosmological surveys to be conducted by
the Multiband Imaging Photometer for SIRTF (MIPS) at 24, 70 and 160
microns, for the guaranteed time observations and the legacy programs,
using the latest knowledge of the instrument. 
In addition to the detector noise and the cirrus confusion noise, we
discuss in detail the derivation of the confusion noise due to
extragalactic sources, that depends strongly on the shape of the
source counts at a given wavelength and on the telescope and detector
pixel sizes. We show that it is wise in general to compare the
classical photometric criterion, used for decades, and the so called
source density criterion to predict the confusion levels. We obtain,
using the model of Lagache, Dole, \& Puget (2002)
limiting fluxes of 50 $\mu$Jy, 3.2 mJy and 36 mJy at 24, 70 and 160
microns respectively. After taking into account other known sources of noise
that will limit the surveys sensitivities, we compute the redshift
distributions of the detected sources at each wavelength, and show
that they extend up to z $\sim$ 2.7 at 24 $\mu m$ and up to z $\sim$
2.5 at 70 and 160 $\mu m$, leading to resolve at most 69, 54 and 24\%
of the Cosmic Infrared Background (CIB) at 24, 70 and 160 microns
respectively. We estimate which galaxy populations will be used to
derive the luminosity function evolution with redshift. 
We also give the redshift distributions of the unresolved sources in
the FIR range, that dominates the fluctuations of the CIB, and a
predicted power spectrum showing the feasibility of fluctuations
(both due to Poissonian and clustered source distributions)
measurements. The main conclusion is that MIPS (and SIRTF in general)
cosmological surveys will greatly improve our understanding of galaxy
evolution by giving data with unprecedented accuracy in the mid and
far infrared range. 
\end{abstract}

%
\section{Introduction}
ISO, the Infrared Space Observatory, performed deep surveys in
the mid (MIR) and far (FIR) infrared range
\cite[for reviews]{genzel2000,franceschini2001}
in order to study galaxy evolution and to constrain the global star
formation rate. Together with other surveys performed from the ground
(e.g. with SCUBA and MAMBO), our view about galaxy evolution in the
infrared, submillimeter and millimeter range became more accurate.

With the information extracted from these cosmological surveys, and in
particular from the source counts, the redshift distribution of the
sources, the spectral energy distribution of the Cosmic Infrared
Background (CIB), and the analysis of the CIB fluctuations, it is
possible to build a coherent view of galaxy evolution and formation in
the infrared and submillimeter range by developing models that fit
all the available data.
Many semi-empirical models exist
\cite[]{roche99,tan99,devriendt2000,dole2000,wang2000,chary2001,franceschini2001,malkan2001,pearson2001,rowan-robinson2001a,rowan-robinson2001b,takeuchi2001,xu2001,wang2002}
and try to address questions about the evolution of infrared galaxies,
inferring the global star formation rate.
These models fit reasonably well the data. Recently, Lagache, Dole \&
Puget (2002) have developed a phenomenological model which satisfies
all the present observational constraints, one of which being the
fluctuations of the background, as a powerful tool to investigate
future observations. 

The availability of new space facilities in the coming years, such as the
Space Infrared Telescope Facility (SIRTF) in early 2003, ASTRO-F, and later in
the decade Planck and Herschel, and on the ground with the Atacama Large
Millimeter Array (ALMA), opens new perspectives to study in detail the
population of infrared galaxies beyond z=1.
Which galaxy populations these facilities will be able to
detect ? What fraction of the CIB will be resolved into sources ? Up
to what redshift will it be possible to construct a luminosity
function and detect any evolution ? What will be the observational
limitations on the cosmological surveys ?

To answer most of these questions prior to any new data being taken,
and to better plan the surveys that will fully use the capabilities of
these new facilities, it is common to use the models to make
predictions, according to today's knowledge. The goal of this paper is
to investigate the properties of the planned SIRTF surveys with the
Multiband Imaging Photometer for SIRTF (e.g. confusion, sensitivity,
redshift distributions), using the \cite{lagache2002}
model as well as the latest knowledge of the MIPS instrument. Detailed
predictions for Herschel, Planck and ALMA are given in \cite{lagache2002}.

The structure of the paper is as follows.
The MIPS instrument and the planned surveys are
described in Sect.~\ref{sect:mipsandgto}.
We discuss the confusion noise due to the galactic cirrus in
Sect.~\ref{sect:cirrus}. 
In Sect.~\ref{sect:model} we summarize the \cite{lagache2002} model,
and one of its applications in 
Sect.~\ref{sect:mapsimus}: the generation of multiwavelength maps.
The general case of the confusion noise due to extragalactic sources is
discussed in Sect.~\ref{sect:confusionnoise}, and the confusion limits for
MIPS are given in Sect.~\ref{sect:confusionlimits}. The total sensitivity
for the surveys is given in Sect.~\ref{sect:sensitivity}. We discuss the
expected results about resolved sources in Sect.~\ref{sect:resolvedsources}
and about the unresolved sources in Sect.~\ref{sect:unresolvedsources}.
%
\section{The MIPS Instrument and the Planned Cosmological Surveys}
\label{sect:mipsandgto}
%
\subsection{MIPS}
\label{sect:mips}
MIPS\footnote{All useful material regarding the SIRTF instruments, including
the characteristics and the simulated beam profiles are available at
the SSC Web site: \url{http://sirtf.caltech.edu/SSC/}}, the
Multiband Imaging Photometer for SIRTF \cite[]{rieke84,young98,heim98}, 
is one of the three SIRTF \cite[]{werner95} focal plane instrument, the others being the
Infrared Camera, IRAC \cite[]{fazio98}, and the 
Infrared Spectrograph, IRS \cite[]{houck95}.
MIPS is composed of three large array detectors, sensitive
at 24, 70, and 160 $\mu m$ respectively. The array sizes 
are 128$^2$, 32$^2$ and $2 \times 20$ pixels respectively, and the
detector material is Si:As BIB, Ge:Ga and stressed Ge:Ga respectively.
Among the main key features of MIPS, there are 1) the large size of the
arrays, 2) the technical achievements in the detectors, 3) the
calibration strategy of the FIR arrays \cite[]{engelbracht2000} with
frequent stimulator flashes tracking the responsivity variations, and
4) the presence of a scan mirror allowing an efficient and redundant
sky coverage of 5 arcminute wide stripes, simultaneously at all three
wavelengths. 

The beam profile characteristics play an important role in computing the confusion
limits; they have been 
generated using the STinyTim software, which is an 
updated version for SIRTF of the TinyTim software for HST
\cite[]{krist93}. Tab.~\ref{tab:mips_psf} summarizes the main
characteristics of the pixels and beam profiles for MIPS.
%
%
\clearpage
\begin{deluxetable}{lccc}
\tablewidth{0pt}
\tablecaption{Some MIPS Instrumental Characteristics: Pixel size, Beam
Profile, Noise\label{tab:mips_psf}}
\tablehead{
\colhead{} &\colhead{24 $\mu m$} & \colhead{70 $\mu m$} & \colhead{160 $\mu m$}
}
\startdata
        pixel size (")                          & 2.55          & 9.84          & 16.0 \\
        FWHM  (")$^a$                           & 5.6           & 16.7          & 35.2 \\
	pixel solid angle (sr)$^b$	& $1.41 \times 10^{-10}$ & $2.30 \times 10^{-9}$ & $5.87 \times 10^{-9}$ \\
        $\int f(\theta,\phi)d\theta d\phi$ (sr)$^c$       & $1.25 \times 10^{-9}$ & $9.98 \times 10^{-9}$  & $4.45 \times 10^{-8}$ \\
        $\int f^2(\theta,\phi)d\theta d\phi$ (sr)$^d$     & $4.27
\times 10^{-10}$ & $3.45 \times 10^{-9}$ & $1.66 \times 10^{-8}$\\[+5pt]\tableline
	1$\sigma_p$ for 10s integration$^e$    		& 0.22 mJy     & 2.0 mJy     & 6.6 mJy \\
\enddata
\tablenotetext{a}{Measured from STinyTim models}
\tablenotetext{b}{Pixel solid angle in sr}
\tablenotetext{c}{Integral of the beam profile $f(\theta,\phi)$}
\tablenotetext{d}{Integral of the squared beam profile $f^2(\theta,\phi)$ (used in
Eq.~\ref{eq:sigma_slim})}
\tablenotetext{e}{1$\sigma$ photon (and instrumental) noise for 10s
integration, both for scan map and photometry modes (Rieke, Private Communication.)}
\end{deluxetable}
%
\subsection{Cosmological Surveys with MIPS}
\label{sect:deepsurveys}

The currently planned Cosmological Surveys with MIPS are mainly
scheduled through two types of programs: the GTOs (Guaranteed Time
Observers) and the Legacy Programs. Deep IRAC observations are also planned
for all programs, but are not discussed in this paper.
The characteristics of all the following surveys  are summarized in
Tab.~\ref{tab:gto_mips}. 

The MIPS GTO program for Cosmological
Surveys\footnote{\url{http://lully.as.arizona.edu}}
is composed of three surveys,
named Shallow, Deep and Ultra-Deep respectively, whose
characteristics are listed in Tab.~\ref{tab:gto_mips}.
The MIPS GTO program also includes
galaxy cluster observations, aimed at mapping lensed background
galaxies. In addition, some IRAC and IRS GTO programs share the same
targets or directly contribute to some of them. 

Two of the six Legacy Programs are focussed on cosmological surveys:
SWIRE\footnote{\url{http://www.ipac.caltech.edu/SWIRE/}}
(SIRTF Wide-area InfraRed Extragalactic Survey) and
GOODS\footnote{\url{http://www.stsci.edu/science/goods/}} 
(Great Observatories Origins Deep Survey).
Schematically, for MIPS observations, the SWIRE surveys have the same
observational strategy as the GTO shallow survey, but will cover a
larger sky area (65 Sq. Deg.), and the GOODS
surveys are similar to the GTO ultra-deep survey but will observe a
0.04 Sq. Deg. field at 24 $\mu m$ with more depth.

Finally, an early survey in the SIRTF mission will be
conducted with MIPS and IRAC to verify the observational strategies:
the First Look Survey
(FLS\footnote{\url{http://sirtf.caltech.edu/SSC/fls/extragal}}) of the
extragalactic component. Given the similarities with other surveys, we
won't discuss this survey specifically. 

%
\clearpage
\begin{deluxetable}{lccccc}
\tablewidth{0pt}
\tablecaption{MIPS Cosmological Survey Key Features \label{tab:gto_mips}}
\tablehead{
\colhead{Survey} &\colhead{MIPS Observation} &\colhead{Area} &\colhead{24 $\mu m$} & \colhead{70 $\mu m$} & \colhead{160 $\mu m$}\\
\colhead{} &\colhead{Mode$^a$} &\colhead{Sq. Deg.$^b$} &\colhead{$t_{int}$ (s)$^c$} & \colhead{$t_{int}$ (s)$^c$} & \colhead{$t_{int}$ (s)$^c$}

}
\startdata
	Shallow$^d$    & Scan Medium (2 passes)  & 9    & 80s  & 80s   & 8s \\
	Deep$^d$       & Scan Slow (12 passes)   & 2.45 (6 $\times$ 0.41) & 1200s & 1200s & 120s  \\	
	Ultra Deep$^d$ & Photometry              & 0.02 & 14700s & 12500s & -- \\
	Clusters$^d$   & Photometry              & 0.2 (28 $\times$ 0.007) & 3300s & 400s & 80s  \\
	SWIRE$^e$      & Scan Medium (2 passes)  & 65 (7 fields) & 80s   & 80s   & 8s    \\
	GOODS$^e$      & Photometry              & 0.04 & 36000s & --    & -- \\	
	FLS$^f$        & Scan Medium (2 passes)  & 5    & 80s    & 80s   & 8s \\
	FLS verif$^f$  & Scan Medium (10 passes) & 0.25 & 400s   & 400s  & 40s \\[+5pt]
\enddata
\tablenotetext{a}{MIPS Observation mode. For the surveys, two modes
are used: photometry, and scan map. In the case of scan maps, the rate
is given: medium (4s/frame) or slow (10s/frame), as well as the number
of passes.}
\tablenotetext{b}{Total surface of the survey. If more than one field,
the detail of the number of fields and the approximate size is also
given.}
\tablenotetext{c}{Integration time in seconds per sky pixel.}
\tablenotetext{d}{GTO program.}
\tablenotetext{e}{Legacy program.}
\tablenotetext{f}{First Look Survey.}
\end{deluxetable}
%
%
\subsection{Sensitivity}
\label{photonnoise}
The noise in the MIPS instrument is the sum of the detector-related noise
(e.g. read noise, linearity correction noise, instantaneous flat field
noise), the cosmic rays, and the photon noise. 
The noise budget is dominated by the photon noise (Rieke, private
communication).
For simplicity, we will call the total noise {\it photon noise} $\sigma_p$, 
even if all the instrumental noise sources are taken into account. 
Tab.~\ref{tab:mips_psf} gives the 1 $\sigma_p$ noise in scan
map mode for a 10s integration (Scan Map mode or Photometry Mode).  
The upper part of Tab.~\ref{tab:total_noise_mips} gives the 1 $\sigma_p$
noise for the different integrations planned for the
surveys. Notice that noise caused by any systematic effect is not
taken into account here. It has been shown however for ISOCAM that the
latter noise source do not degrade the final sensitivity
\cite[]{miville-deschenes2000}. 

%
\section{Cirrus Confusion Noise}
\label{sect:cirrus}
Previous works \cite[]{helou90,gautier92,kiss2001} studied in detail
the confusion noise due to Galactic cirrus $\sigma_{gc}$, and showed
that in most cases it can be simply parametrized as follows:
\begin{equation}
\rm 
\sigma_{gc} = 0.3 {\times}  (\lambda_{100} )^{2.5}
   ( D_m ) ^{-2.5} {\langle B_{\lambda} \rangle} ^{1.5}
\label{NHB}
\end{equation}
where $\sigma_{gc}$ is in mJy, $\lambda_{100}$ is the wavelength ratio
${\lambda}\over{100\,{\mu}m}$, $D_m$ is the telescope diameter in m,
and $\langle B_{\lambda} \rangle$ is the brightness in MJy/sr
\cite[]{helou90}. 
\cite{kiss2001} report that this parameterization underestimates $\sigma_{gc}$
by a factor of 2. However, their estimate of $\sigma_{gc}$ includes a
contribution from CIB fluctuations which is known to be significant
\cite[]{lagache2000a}, and so we can use the parameterization when
we are only concerned with the Galactic cirrus component 

Using Fig.~1 of \cite{boulanger2000a} for the spectrum
of the diffuse ISM, we extrapolate the mean brightness at 100 $\mu m$
$\langle B_{100} \rangle$ of 0.5 MJy.sr$^{-1}$ (corresponding to an HI
column density of $10^{20} cm^{-2}$, typical for cosmological surveys)
at 24, 70 and 160 $\mu m$.
We then derive the corresponding cirrus confusion noise $\sigma_{gc}$ from
Eq.~\ref{NHB}. The results are given in Tab.\ref{tab:cirrus_confusion}.
For most of the cosmological fields, where the cirrus brightness $\langle
B_{100} \rangle$ is less than 1 MJy.sr$^{-1}$, the cirrus confusion noise is often
negligible or is a minor contribution to the total noise.
In this work, we will thus only consider the confusion due to extragalactic
sources, letting the reader adding the cirrus confusion noise
appropriate to its own purpose.
%
\clearpage
\begin{deluxetable}{lccc}
\tablewidth{0pt}
\tablecaption{Cirrus Confusion Noise\label{tab:cirrus_confusion}}
\tablehead{
\colhead{} &\colhead{24 $\mu m$} & \colhead{70 $\mu m$} & \colhead{160 $\mu m$}
}
\startdata
	$\langle B_{\lambda} \rangle$ MJy.sr$^{-1}$$^a$& 0.03        & 0.12        & 1.5  \\
	$\sigma_{gc}^b$	               & 0.06 $\mu$Jy & 7.6 $\mu$Jy  & 2.7 mJy \\[+5pt]	
\enddata
\tablenotetext{a}{Cirrus brightness for MIPS bands; this cirrus has a brightness $\langle B_{100}
\rangle = 0.56$ MJy.sr$^{-1}$ at 100 $\mu m$, corresponding to $N_{HI} = 10^{20}
cm^{-2}$. We used the dust spectrum from \cite{boulanger2000a}.}
\tablenotetext{b}{1 $\sigma_{gc}$ Cirrus Confusion Noise derived from Eq.~\ref{NHB}
\cite[]{helou90}}
\end{deluxetable}
%
\section{Model of Infrared Galaxy Evolution}
\label{sect:model}
In addition to the photon noise and cirrus confusion noise, the noise
due to the extragalactic sources is certainly the dominant noise for
the cosmological surveys. The Lagache, Dole \& Puget (2002) model is
used to describe this component.

This model fits, besides the CIB intensity, source counts, the redshift
distribution and colors, and the additional observational constraint of the
CIB fluctuations. It describes only the dust emission part of the galaxies
in the 4 $\mu m$ to 1.5 mm wavelength range. It is a phenomenological
model based on two galaxy populations: the IR emission of normal
spirals where optical output dominates and a starburst
population. Each population is characterized by an SED family and an evolving
luminosity function, described by a small number of parameters.
The predictions of this model thus cover well the observed wavelength
range from 8 $\mu m$ to 3 mm. It does not include source clustering. The
confusion is computed for the Poisson contribution, and the clustering
might slightly change the confusion limits; this will be investigated
in forthcoming papers \cite[]{blaizot2003,sorel2003}.

The model outputs as well as some programs are publicly available on
our web
pages\footnote{\url{http://www.ias.fr/PPERSO/glagache/act/gal\_model.html} and
\url{http://lully.as.arizona.edu/Model/}}.

%
\section{Simulating the Multi-Wavelength IR Sky}
\label{sect:mapsimus}
One of the applications of the model of \cite{lagache2002} to plan future
observations, is the creation of simulated maps of the infrared and
submillimeter sky. The main purposes of the simulations are:
1) to test the calibration and map making algorithms, 2) to test and
validate the source extraction and photometry procedures, check the
completeness, and 3) to test other algorithms, such as HIRES or band
merging procedures, to improve source detections in the FIR
range. Results of these simulations will be the subject of a
forthcoming paper. 

The maps\footnote{Images of the maps are available on our web
site \url{http://lully.as.arizona.edu/Simulations/}}, available for
public use upon request, are sampled with 2" pixels and have sizes
ranging from $1024^2$ to $4096^2$ (0.32 to 5 Sq. Deg.). 
The simulated maps contain three components: an extragalactic
component (IR galaxies), a galactic foreground component (cirrus), and
a zodiacal light component. The following is a brief description of
each component. 

The \cite{lagache2002} model evolving luminosity functions are used to create the
extragalactic component in simulated maps over a wide range of
wavelengths relevant to today and future studies (mainly for ISO,
SIRTF, ASTRO-F, Planck, Herschel, SCUBA, MAMBO, ALMA). For computational
efficiency, we add in the maps sources only in the redshift range 0 to 5.

The galactic foreground component, the cirrus, is build as follows: the
spatial structure is taken from an actual 100 $\mu m$ cirrus in the
IRAS recalibrated maps of \cite{schlegel98}, and the scale extrapolation
to smaller scales uses the properties of the cirrus power spectrum
from \cite{gautier92}. We then use the cirrus spectrum of \cite{boulanger2000}
to compute this component at other wavelengths.

The zodiacal light component is a constant value in our maps, taken
from Tab.4 of \cite{kelsall98} for high ecliptic and galactic latitude
fields.
%
\section{Deriving the Confusion Noise due to Extragalactic Sources}
\label{sect:confusionnoise}
Numerous authors
\cite[]{condon74,hacking87,hacking91,franceschini89,franceschini91,vaisanen2001}
have described the effect of the fluctuations due to the presence of
point sources in a beam. For technological reasons limiting the
telescope diameter compared to the wavelength, these fluctuations play
an important (if not dominant) role in the measurements noise budget
in the mid- and far-infrared, submillimeter and centimeter range for
extragalactic surveys.

Through the rest of the paper, we will use the term {\it confusion limit}
for {\it confusion limit due to extragalactic sources}.
There are two different criteria to derive the confusion noise.
The widely-used {\bf photometric criterion}
(Sect.~\ref{photometriccriterion}) is derived from the fluctuations of
the signal due to the sources below the detection threshold $S_{lim}$
in the beam; it was well adapted for the first generations of space IR
telescopes (IRAS, COBE, ISO).
The {\bf source density criterion}
(Sect.~\ref{sourcedensitycriterion}) is derived from a completeness
criterion and evaluates the density of the sources detected above 
the detection threshold $S_{lim}$, such that only a small fraction of
sources is missed because they cannot be separated from their nearest
neighbor.

We will show that with SIRTF (or other planned telescopes), we need in
general (regardless of the model used) to compare the confusion noise
given by the two criteria, in order not to artificially underestimate
the derived confusion noise. In the frame of the \cite{lagache2002}
model, we'll give our estimates for the confusion.

%
\begin{figure}
\plotone{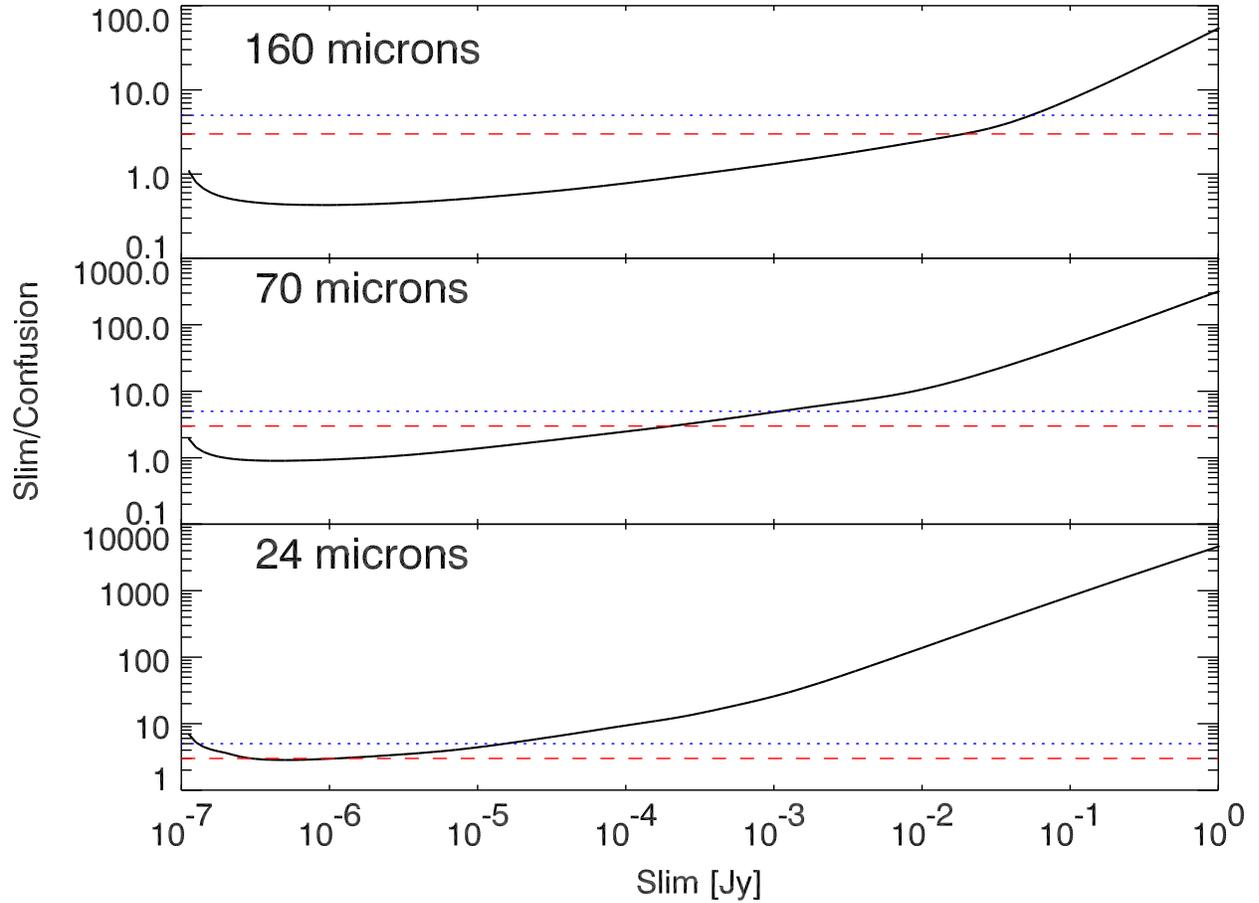}
\caption{\label{fig:plot_sn_conf} Signal to confusion noise
ratio as a function of $S_{lim}$ at 24, 70 and 160 $\mu m$
(solid line). $S_{lim}/\sigma_c=3$ (dash) and $S_{lim}/\sigma_c=5$ (dot) are also plotted. At 24
$\mu m$, $S_{lim}/\sigma_c$ is always greater than 3; using the photometric
criterion for deriving the confusion noise thus leads to a severe
underestimation.}
\end{figure}
%
\subsection{Confusion Noise: General Case}
\label{confusionnoisegeneralsect}
At a given frequency $\nu$ (hereafter the subscript $\nu$ will be omitted), let
$f(\theta,\phi)$ be the two-dimensional beam profile (peak normalized to unity);
let $S$ be the source flux density (hereafter flux) in Jy;
let $dN/dS$ be the differential source counts in $Jy^{-1} sr^{-1}$.\\
The amplitude of the response $x$ due to a source of flux S at
location ${\theta,\phi}$ within the beam is
\begin{equation}
\label{eq:x}
x = S f(\theta,\phi)
\end{equation}
The mean number of responses, $R(x)$ with amplitudes between $x$ and $x + dx$,
from sources present in the beam element $d\Omega$ at position $(\theta, \phi)$
(where $d\Omega = 2 \pi \theta d\theta d\phi$):
\begin{equation}
R(x)dx = \int_{\Omega} \frac{dN}{dS}dSd\Omega
\end{equation}
The total variance $\sigma^2_{c}$ of a
measurement within the beam due to
extragalactic sources of fluxes less than $S_{lim}$ is given by:
\begin{equation}
\sigma_{c}^2 = \int_0^{x_{lim}} x^2 R(x) dx
\end{equation}
where $x_{lim} = S_{lim} \times f(\theta,\phi)$ is the cutoff
response at high flux. This can be rewritten as:
\begin{equation}
\label{eq:sigma_slim}
\sigma_{c}^2 = \int f^2(\theta,\phi)d\theta d\phi \int_{0}^{S_{lim}}S^2 \frac{dN}{dS}dS
\end{equation}
We call $\sigma_{c}$ the {\it confusion noise}, and $S_{lim}$ the {\it
confusion limit}. There are different ways of deriving $S_{lim}$,
and they will be investigated in Sect.~\ref{photometriccriterion} and
\ref{sourcedensitycriterion}. Note that using Eq.~\ref{eq:sigma_slim}
to determine the confusion limit is an approximation. A first
refinement would be to use the limiting deflection $x_{lim}$ rather
than $S_{lim}$, as explained by e.g. \cite{condon74}, and then
introducing the effective beam. For MIPS, this changes the confusion
level by less than 10\%. Nevertheless, this refinement is not enough
since it does not take into account other important parameters,
related to the observational strategy and the analysis scheme, like
the sky sampling, the pixelization (or PSF sampling), and the
source extraction process, that also impact the confusion limit.
Only complete realistic simulations would allow to predict accurately the
confusion level; this next step will be addressed in a forthcoming
paper using our simulations (Sect.~\ref{sect:mapsimus}). The method
presented here aims at providing a theoretical prediction, which can
be considered as a lower limit.  
%
\subsection{Beam Profiles}
\label{beamprofiles}
Before we obtain measurements of the telescope PSF (Point Spread Function) in
flight, we need to use models of the beam
profiles for the predictions of the confusion noise. 
A popular approximation is to use a Gaussian profile with the same
FWHM than the expected PSF, although for SIRTF an Airy function should
be more appropriate. The Gaussian profile is useful for analytical
derivations of the confusion level as a function of the beam size
\cite[]{vaisanen2001}.
We want here to address the question of accuracy using the Gaussian
approximation, the Airy approximation, or the modeled profile.

We compare the integral of the Gaussian profile (as written in
Eq.~\ref{eq:sigma_slim}) with the simulated profile obtained by STinyTim 
(Sect.~\ref{sect:mips}): this leads to a small error in the first 
integral in Eq.~\ref{eq:sigma_slim} at the order of 2 to 10\%
depending on the MIPS wavelength; the difference is larger on
the integral of the profile, about 30\%.
The Gaussian profile is thus a good approximation for computing
analytically the confusion noise, but not for source extraction simulations.

Using an Airy profile gives better results for the profile integral,
with a difference of less than 20\%; the difference on the profile
integrated according to Eq.~\ref{eq:sigma_slim} is worse, at the order
of 10 to 35\%. The Airy profile is 
thus better suited for source extraction simulations than for
confusion noise estimates. 

The use of the simulated STinyTim profiles (see
Tab.~\ref{tab:mips_psf}) is at present our best approximation
of the flight profiles. Indeed, \cite{lagache2001} have shown in the
case of ISOPHOT that the theoretical profile is in good agreement with
the actual profile.
%
\subsection{The Photometric Criterion for Confusion Noise}
\label{photometriccriterion}

The {\bf photometric criterion} is defined by choosing the signal to
noise ratio $q$ between the faintest source of flux $S_{lim}$, and the RMS noise
$\sigma_{c}$ due to fluctuations from beam to beam (due to sources fainter than
$S_{lim}$), as described in Eq.~\ref{eq_q}: 
\begin{equation}
\label{eq_q}
q = \frac{S_{lim}}{\sigma_{c\, phot}(S_{lim})}
\end{equation}
$S_{lim}$, and thus $\sigma_{c}$, is found by solving Eq.~\ref{eq_q} through an iterative
procedure.
$q$ is usually chosen with values between 3 to 5, depending on the
objectives followed. Notice that $\sigma_{c\,
phot}$ increases with $q$, as given in
the upper part of Tab.~\ref{tab:confusionmips}. As a guideline, if one assumes a power law
for the shape of the differential source counts ($\frac{dN}{dS}
\propto S^{\alpha}$, with $\mid \alpha\, | < 3$), then $\sigma_{c\, phot}$ varies with $q$ like
$\sigma_{c\, phot} \propto q^{-\frac{3 + \alpha}{1 + \alpha}}$. This
can be used in the Euclidean regime ($\mid \alpha\, | = 2.5$). Note that
$\alpha$ has the same meaning as $-\gamma$ in \cite{condon74}.

To illustrate the behavior of the implicit Eq.~\ref{eq_q},
Fig.~\ref{fig:plot_sn_conf} gives $S_{lim} / \sigma_{c phot}$ as a 
function of $S_{lim}$ given by Eq.~\ref{eq:sigma_slim}, as well as the
constant ratio $q = S_{lim} / \sigma_{c phot}$ for $q=$ 3 and 5.
This plot illustrates that using $q=3$ at 24 $\mu m$ does not give a
well defined solution, as the $S_{lim} = 3 \times \sigma_{c \, phot}$
line is almost tangent to the curve $\sigma_{c \, phot}(S_{lim})$; in
this case, the signal to photometric confusion noise is always greater
than three. 
%
\subsection{The Source Density Criterion for Confusion Noise}
\label{sourcedensitycriterion}
A second criterion for the confusion can be defined by setting the
minimum completeness of the detection of sources above $S_{lim}$,
which is driven by the fraction of sources lost in the detection process
because the nearest neighbor with flux above $S_{lim}$ is too close
to be separated\footnote{The completeness is also affected by the noise which
modifies the shape of the source counts: the so-called
Malmquist-Eddington bias. For the sake of simplicity, this bias was
not taken into account}.
For a given source density $N$ (Poisson distribution) corresponding to sources with fluxes above $S_{lim}$, the
probability P to have the nearest source of flux greater or equal to $S_{lim}$ located
closer than the distance $\theta_{min}$ is:
\begin{equation}
\label{eq:p}
P(<\theta_{min}) = 1 - e^{- \pi N \theta_{min}^2}
\end{equation}
Using $\theta_{FW}$, the Full Width at Half Maximum of the beam profile and $k$, we
parameterize $\theta_{min}$ as:
\begin{equation}
\label{eq:l_F_k}
\theta_{min} = k \times \theta_{FW}
\end{equation}
Fixing a value of the probability $P$ gives a corresponding density of
sources, $N_{SDC}$ (SDC stands for Source Density Criterion):
\begin{equation}
\label{eq:n_p}
N_{SDC} = - \frac{log(1 - P(<\theta_{min}))}{\pi k^2  \theta_{FW}^2}
\end{equation}
At a given wavelength, there is a one-to-one relationship between the
source density and the flux, given by the source counts; thus
$S_{SDC}$ is determined with $N_{SDC}$ with our source count model.
We identify $S_{SDC}$ as $S_{lim}$, and can then
compute the confusion noise using the source density criterion
$\sigma_{SDC}$ using Eq.~\ref{eq:sigma_slim}, as a function of P and k. 

We define the {\bf source density criterion} for deriving the
confusion noise, by choosing a value of $P(<\theta_{min})$ and $k$,
the latter can be determined e.g. by simulations.
$S_{lim}$ is the limiting flux, such as there is a chosen
probability $P(<\theta_{min})$ of having two sources of flux above
$S_{lim}$ at a distance of less than $\theta_{min}= k \, \theta_{FW}$.
 
We made simulations of source extraction with DAOPHOT and checked 
that $k=0.8$ is an achievable value; this is also in agreement with
the results from \cite{rieke95}.
We thus use $k=0.8$. We use $P=10$\%, meaning that 10\%
of the sources are too close to another source to be extracted.
The corresponding source density is,
as explained in Tab.~1 of \cite{lagache2002}\footnote{
Using the relation, valid for both Airy and Gaussian profiles, linking
$\theta_{FW}$, the Full Width at Half Maximum of the beam profile, and
$\Omega$, the integral of the beam profile: 
$\Omega \simeq 1.14 \, \theta_{FW}^2$ \cite[]{lagache2002}.},
$1/16.7 \, \Omega$. 
The middle part of Tab.~\ref{tab:confusionmips} gives $S_{lim\,
sdc}$ using the Source Density Criterion, and the corresponding
equivalent $q_{sdc}$ which is the ratio $S_{lim}/\sigma_{sdc}$.

One the one hand, the Photometric and Source Density criteria give
almost identical results in the simple Euclidean case, if one takes
$q=3$, $k=1$, and a maximum probability to miss a source too close to
another one of 10\%. In this classical case, confusion
becomes important for a source density corresponding to one source per
30 independent instrumental beams. 
On the other hand, when the relevant LogN-LogS function departs
strongly from Euclidean, the two criteria give very different
results for these reasonable values of $q$, $k$, and $P$.
Furthermore, for specific astrophysical problems, one might want to
choose significantly different values of these parameters. In that
case, the two criteria might not be equivalent.
For instance at 70 $\mu m$, increasing  P to 20\%, 45\% and 60\%
respectively (instead of the 10\% we're using), give a confusion limit
identical to q=5, 4 and 3 respectively, even if in the last case 60\%
of the sources will be missed.

%
\section{Confusion Limits for MIPS and Comparison with Other Works}
\label{sect:confusionlimits}
%
\subsection{Confusion Limits for MIPS}
\label{confusionlimitsformips}
Comparing the photometric (Sect.~\ref{photometriccriterion}) and the
source density (Sect.~\ref{sourcedensitycriterion}) criteria for the
confusion, we conclude that for MIPS, the source density criterion is
always met before (i.e. at higher flux) the photometric criterion
using $q \simeq 4$. At 160 $\mu m$, the two criteria become
identical. \cite{lagache2002} show that for all the IR/submm space
telescopes of the coming decade, the break point between the two
criteria is at around 200 $\mu m$.

SIRTF, together with its high sensitivity and its well sampled PSFs,
will probe a regime in the source counts where the classical
photometric criterion is no longer valid. The main reasons are 
1) the steep shape of the source counts, and 2) the fact that a
significant part of the CIB will be resolved into sources
(Sect.~\ref{resolvingcib}). This leads to a high source density at faint
detectable flux levels, that actually limits the ability to detect
fainter sources. In this case, the limiting factor is not the
fluctuations of the sources below the detection limit (photometric
criterion) but the high source density above the detection limit
(source density criterion). 

For SIRTF, we thus use the {\it source density criterion} for deriving
the confusion noise and limit.

For the previous generations of infrared telescopes
(IRAS, ISO), it is interesting to compare the two criteria, and
usually they converge to the same answer -- a direct consequence of 
undersampling a large PSF that doesn't allow to probe deeper the
source counts. In this case, the photometric criterion is applicable
and has been widely used.

The confusion noise and the confusion limit for MIPS are given in
the lower part of Tab.~\ref{tab:confusionmips}.

\begin{figure}
\plotone{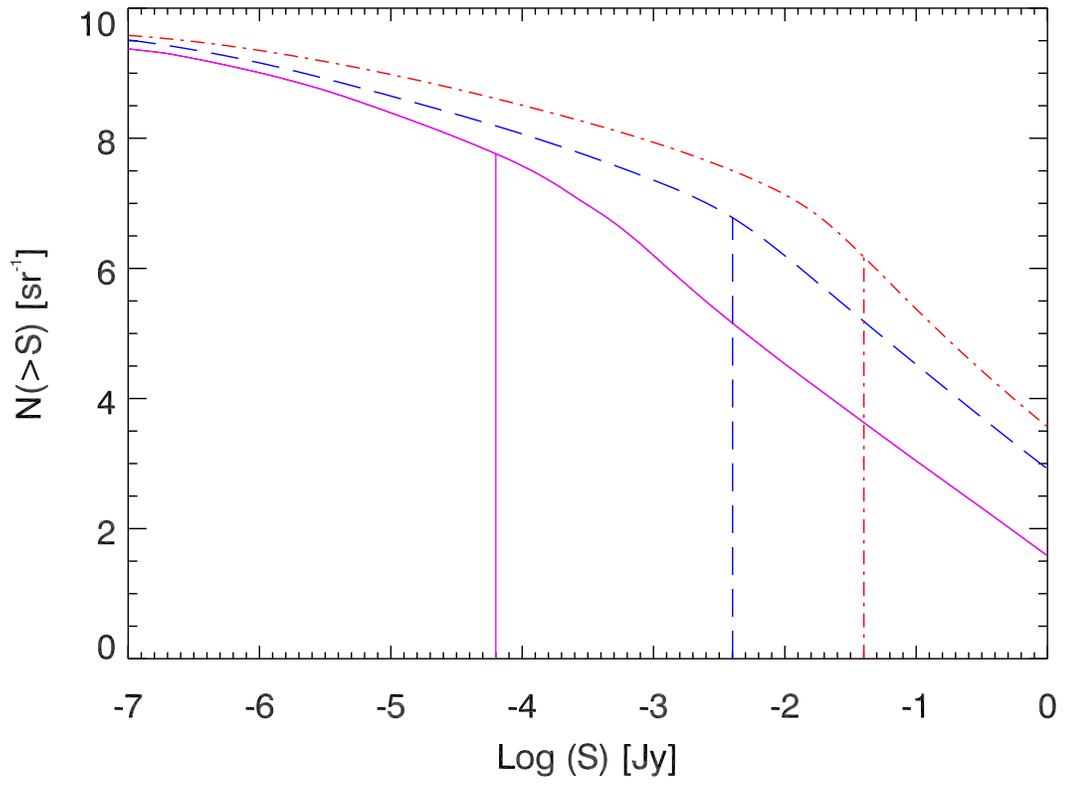}
\caption{\label{fig:counts_multiwv} Integral Source Counts from
our model at 24 (line), 70 (dash) and 160 $\mu m$ (dot-dash), and
Confusion Limits $S_{lim}$ from Tab.~\ref{tab:confusionmips}.}
\end{figure}

Fig.~\ref{fig:counts_multiwv} represents the integral source counts at
24, 70 and 160 $\mu m$ respectively. At these wavelengths, the
confusion limits, given in Tab.~\ref{tab:confusionmips}, correspond to
source densities of $6.9 \times 10^7$, $7.8 \times 10^6$, and $1.9 \times 10^6$
per steradian at 24, 70 and 160 $\mu m$ respectively. 
This corresponds to 11.5, 12.8 and 12.0 beams per sources at
24, 70 and 160 $\mu m$ respectively. The derived values are slightly
lower than the ``generic'' case discussed in
Sect.~\ref{sourcedensitycriterion} of 16.7 sources per beam, the
difference coming from the use of a simulated 
beam profile rather than a Gaussian profile.

%
\subsection{ISO at 170 $\mu m$}
\label{iso170}
The data of the 4 Sq. Deg. FIRBACK survey \cite[]{dole2001} performed
with ISO at 170 $\mu m$ allowed to directly measure the sky confusion
level. This provides a rare opportunity to test the model.

The confusion level was measured at 170 $\mu m$ at 1$\sigma_c = 45$
mJy, and the 4$\sigma_c$ limit (180 mJy) corresponds to 52 beams per
source \cite[]{dole2001}. 
 
Using our model with the actual PSF \cite[]{lagache2001} and the
photometric criterion (valid in this case), we obtain 1$\sigma_c= 40$ mJy,
and for $q=4$, $S_{lim}=158$ mJy; this flux limit corresponds to 40
beams per source. 

The good agreement comforts the quality of the model for estimating
the confusion level from modeled source counts.

%
\clearpage
%
\begin{deluxetable}{lccc}
\tablewidth{0pt}
\tablecaption{Confusion Limits with different Criteria, and Final
Confusion Limits\label{tab:confusionmips}}
\tablehead{
\colhead{} &\colhead{24 $\mu m$} & \colhead{70 $\mu m$} & \colhead{160 $\mu m$}
}
\startdata
\multicolumn{4}{l}{$S_{lim}$ and $q$ using the Photometric Criterion$^a$}\\ \tableline
	$S_{lim}$, $q=3$ 	& -- 		& 0.20 mJy 	& 20 mJy  \\
	$S_{lim}$, $q=4$ 	& 7.1 $\mu$Jy 	& 0.56 mJy 	& 40 mJy  \\
	$S_{lim}$, $q=5$ 	& 15.8 $\mu$Jy 	& 1.12 mJy 	& 56 mJy \\[+5pt]
\tableline
\multicolumn{4}{l}{$S_{lim}$ and $q$ using the Source Density Criterion$^b$}\\ \tableline
	$S_{lim}$ 		& 50 $\mu$Jy 	& 3.2 mJy 	& 36 mJy \\
	$q_{sdc}$ 		& 7.3 		& 6.8 		& 3.8  \\[+5pt]
\tableline
\multicolumn{4}{l}{{\bf $S_{lim}$ and $q$ using the Best Estimator}$^c$}\\ \tableline
	{\bf $S_{lim}$} & {\bf 50 $\mu$Jy} 	& {\bf 3.2 mJy} & {\bf 36.0 mJy}\\
	$q$ 			& 7.3$^d$ 	& 6.8$^d$ 	& 3.8$^{d,e}$  \\[+5pt]
\enddata
\tablenotetext{a}{$S_{lim}$ using the Photometric
Criterion, for different values of $q$}
\tablenotetext{b}{$S_{lim}$ using the Source Density 
Criterion, and the equivalent values of $q_{sdc}$.}
\tablenotetext{c}{$S_{lim}$ and $q$ of the best confusion
estimator. These values are our confusion limits.}
\tablenotetext{d}{using the Source Density Criterion}
\tablenotetext{e}{in this case, the Photometric and Source Density Criteria agree.}

\end{deluxetable}
%

%
\begin{figure}
\plotone{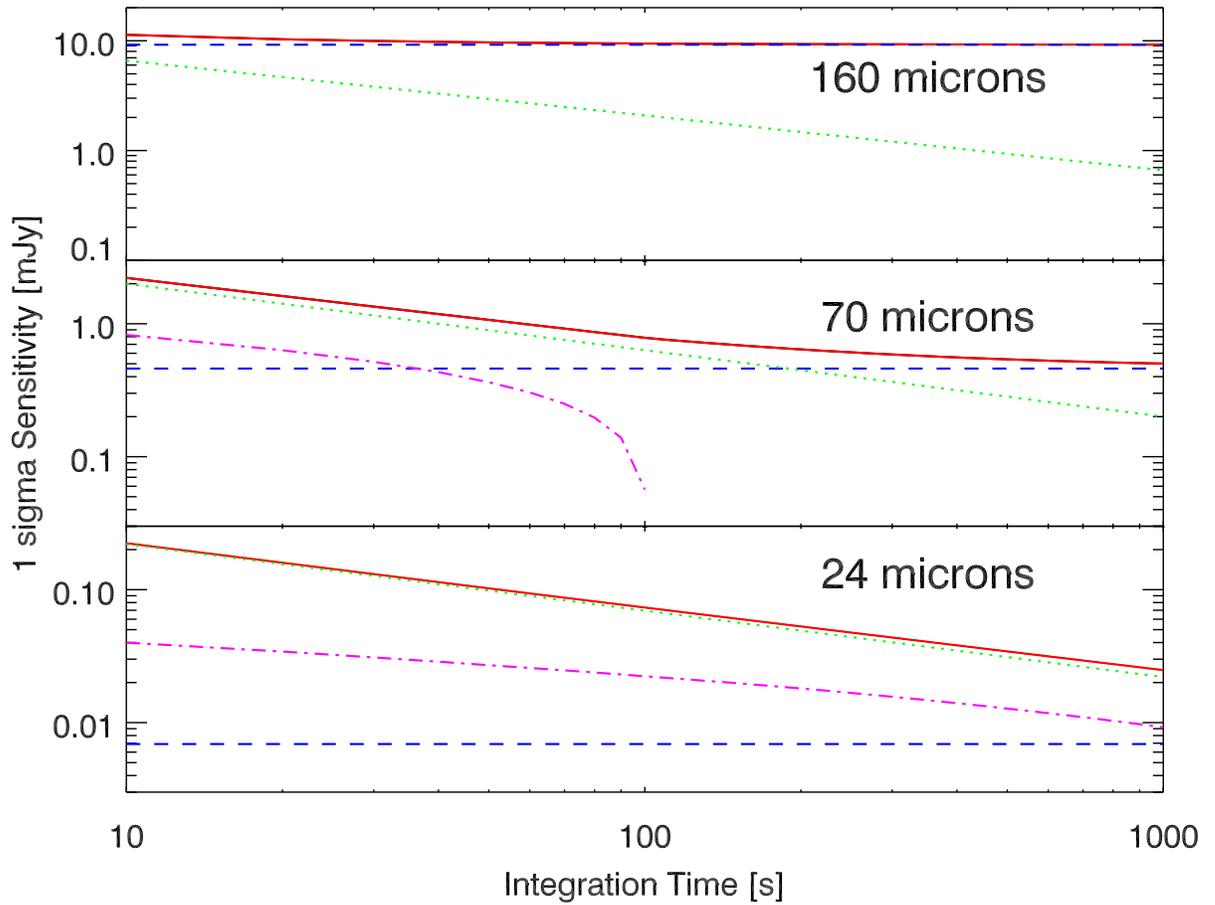}
\caption{\label{fig:total_noise_all} $1 \sigma$ sensitivity of the
scan maps as a function of integration time at 24, 70 and 160 $\mu
m$. Solid Line: Total $1 \sigma$ Sensitivity; Dash: Confusion level $\sigma_{c}$ ;
Dot: photon noise; Dash-dot: additional confusion noise.}
\end{figure}
%
\subsection{Comparison with Other Determinations}
\cite{xu2001} computed the confusion limit $S_{lim}$ with the
photometric criterion using $q=3$ for MIPS and get 33 $\mu$Jy, 3.9 mJy and
57 mJy at 24, 70 and 160 $\mu m$ respectively. This corresponds
respectively to 8, 17 and 31 beams per source. Our estimates
are thus compatible at 70 $\mu m$, but slightly different at 24 and 160
$\mu m$. Their use of the photometric criterion at 24 $\mu m$
significantly underestimates the confusion level. At 160, their redshift
distribution seems to overestimate (at the FIRBACK flux limits) the
population peaking at z $\sim$ 1 \cite[]{patris2002}, which may
suggest a difference in the $dN/dS$ distribution, that directly
affects the predicted confusion levels.

\cite{franceschini2002}, based on the model of
\cite{franceschini2001}, give preliminary $5 \sigma_c$ confusion limits 
for MIPS at 24, 70 and 160 $\mu m$ and get 85 $\mu$Jy, 3.7 and 36 mJy
respectively. This corresponds respectively to 19, 15 and 12 beams per
source. The values for the far infrared are in good agreement with our predictions. 
However, a more refined comparison needs to be done when details of
their computation will be published, especially in the mid infrared.

Other models exist
\cite[]{roche99,tan99,devriendt2000,wang2000,chary2001,pearson2001,rowan-robinson2001b,takeuchi2001,wang2002},
but do not specifically address the point of predicting the confusion
limits for SIRTF. \cite{malkan2001} and \cite{rowan-robinson2001a} make
predictions. The former use, as a photometric criterion, 1 source per
beam. The latter uses 1 source per 40 beams, leading to $S_{lim}$ of
135 $\mu$Jy, 4.7 mJy and 59 mJy at 24, 70 and 160 $\mu m$ respectively.

%
\subsection{The 8 $\mu m$ Case}
Our model reaches its limit around 8 $\mu m$ because our
SEDs are not designed for wavelengths shorter than $4 \mu m$. However,
it fits all observables at wavelengths longer than 7 $\mu$m.
We can thus predict the confusion level. As for the 24 $\mu m$, the
confusion level will be low, and will not limit the extragalactic
surveys. 

At 8 $\mu m$, the photometric criterion does not provide a meaningful
confusion limit because the $S_{lim}/\sigma_{c}$ ratio is always greater than 10.
We obtain, using the source density criterion, $S_{lim} = 0.45
\,\mu$Jy and $\sigma_{c} = 0.05 \,\mu$Jy, leading to $q=9.72$. 

The values from \cite{vaisanen2001} are $S_{lim} = 3 - 4 \,\mu$Jy, 
$\sigma_{c} = 0.40 - 0.51 \,\mu$Jy and $q=10.0$.
Our estimation of the confusion level for this IRAC band is lower by
a factor of $\sim 7$. This discrepancy comes in fact from the
source counts themselves: we underpredict the source density by a
factor 7 to 8 in the range at 0.1 to 1 $\mu$Jy, even if both models
reproduce the ISO counts. This is expected from a model which accounts
properly for the dust emission but does not model the stellar emission
of high redshift galaxies.
When using the
counts from \cite{vaisanen2001}, we agree with their published
values. \cite{vaisanen2001}, in their Sect.~5.3, discuss the
sensitivity of the predicted confusion levels to the shape of the
source counts, and the constraints of the modeled source counts by
the data. Their conclusion is that, although the 7 $\mu m$ ISOCAM source
counts above 50 $\mu$Jy agree within uncertainties, the models below
1 $\mu$Jy are not much constrained. As a result,
the predictions for the confusion level down to the IRAC sensitivity
can be as different as a factor of 10. We confirm this analysis.

%
\section{Sensitivity in the MIPS Final Maps}
\label{sect:sensitivity}
In this section we compute the sensitivity of the MIPS surveys
as a function of the integration time. The total noise $\sigma_{tot}$ is
\cite[]{lagache2002}:
\begin{equation}
\label{eq:sigmatot}
\sigma_{tot} = \sqrt{\sigma_{p}^2 + \sigma_{c}^2 + \sigma_{add}^2 }
\end{equation}
where $\sigma_{p}$ is the photon noise (Sect.~\ref{photonnoise}), $\sigma_{c}$ is
the confusion noise (Sect.~\ref{sect:confusionlimits} \& Tab.~\ref{tab:confusionmips}), and $\sigma_{add}$
is the additional confusion noise. This additional confusion noise is
only present when the photon noise exceeds the confusion noise: in this
case, $\sigma_{add}$ accounts for the confusion due to bright sources above the
confusion limit but below the photon noise. $\sigma_{add}$ is
computed, when $5 \sigma_{p} > S_{lim}$:
\begin{equation}
\label{eq:sigmaadd}
\sigma_{add}^2 = \int f^2(\theta,\phi)d\theta d\phi \int_{S_{lim}}^{5 \sigma_{p}}S^2 \frac{dN}{dS}dS
\end{equation}
Fig.~\ref{fig:total_noise_all} shows $\sigma_{tot}$ and the relative
contributions of $\sigma_{p}$, $\sigma_{c}$ and $\sigma_{add}$ as a
function of the integration time.
It appears that the 160 $\mu m$ data are confusion limited even with
short integrations.
At 70 $\mu m$, the
confusion should dominate the noise for exposures longer than 100s,
and $\sigma_{add}$ is a small component in the first 50s and is
negligible after.
At 24 $\mu m$ we do not expect the data to be confusion limited, and 
$\sigma_{add}$ is between 5 and 3 times smaller than the photon noise.

%
\clearpage
\begin{deluxetable}{lccc}
\tablewidth{0pt}
\tablecaption{Sensitivities of the planned Cosmological Surveys: 1 $\sigma_{p}$ (photon
noise only), 1 $\sigma_{tot}$, and final sensitivity (see text).\label{tab:total_noise_mips}}
\tablehead{
\colhead{} &\colhead{24 $\mu m$} & \colhead{70 $\mu m$} & \colhead{160 $\mu m$}
}
\startdata
\multicolumn{4}{l}{1 $\sigma_{p}$ Sensitivity (does not include sky confusion)$^a$}\\ \tableline
	Shallow 1$\sigma_p$    & 78 $\mu$Jy  & 0.71 mJy    & 6.6 mJy  \\
	Deep 1$\sigma_p$       & 20 $\mu$Jy  & 0.18 mJy    & 1.9 mJy  \\
	Ultra Deep 1$\sigma_p$ &  6 $\mu$Jy  & 0.06 mJy    & -- \\
	Clusters 1$\sigma_p$   & 12 $\mu$Jy  & 0.32 mJy    & 2.3 mJy \\
	SWIRE 1$\sigma_p$      & 78 $\mu$Jy  & 0.71 mJy    & 6.6 mJy  \\
	GOODS 1$\sigma_p$      &  4 $\mu$Jy  & --    & -- \\	
\tableline
\multicolumn{4}{l}{1 $\sigma_{tot}$ Final Sensitivity of the Surveys$^b$}\\ \tableline
	Shallow 1 $\sigma_{tot}$   & 82  $\mu$Jy  & 0.87 mJy  & 11.3 mJy \\
	Deep 1 $\sigma_{tot}$      & 23  $\mu$Jy  & 0.49 mJy  & 9.4 mJy \\
	Ultra Deep 1 $\sigma_{tot}$&  9  $\mu$Jy  & 0.46 mJy  & -- \\
	Clusters 1 $\sigma_{tot}$  & 15  $\mu$Jy  & 0.55 mJy  & 9.5 mJy \\
	SWIRE 1 $\sigma_{tot}$     & 82  $\mu$Jy  & 0.87 mJy  & 11.3 mJy \\
	GOODS 1 $\sigma_{tot}$     &  8  $\mu$Jy  & --    & -- \\
\tableline
\multicolumn{4}{l}{{\bf Final sensitivities of the Surveys}$^c$}\\ \tableline
	Shallow    & 392  $\mu$Jy  & 4.7 mJy   & 48 mJy \\
	Deep       & 112  $\mu$Jy  & 3.2 mJy   & 36 mJy \\
	Ultra Deep & 59   $\mu$Jy  & 3.1 mJy   & -- \\
	Clusters$^d$& 79   $\mu$Jy  & 3.5 mJy   & 37 mJy \\
	SWIRE      & 392  $\mu$Jy  & 4.7 mJy   & 48 mJy \\
	GOODS      & 54   $\mu$Jy  & --   & -- \\	
\enddata
\tablenotetext{a}{1 $\sigma_p$ sensitivities; just includes photon noise.}
\tablenotetext{b}{1 $\sigma_{tot}$ sensitivities, given as a {\it
guideline}; it includes the confusion, the photon (instrumental) and
the additional confusion noise components. Notice 
that it is {\it incorrect} to take $5 \sigma_{tot}$ as a confusion level for
surveys (see text).}
\tablenotetext{c}{{\bf Final sensitivities} (see text) of the planned
Cosmological Surveys. Includes in a proper manner the confusion noise
and photon noise.}
\tablenotetext{d}{the given sensitivities do not take into account
here the properties of background lensed galaxies.}
\end{deluxetable}

The middle part of Tab.~\ref{tab:total_noise_mips} gives the $1 \sigma_{tot}$
sensitivity for the surveys, and includes the confusion, the
instrumental and the additional confusion noise components. Notice
that these $1 \sigma_{tot}$ values are given as a guideline, knowing that taking $5
\sigma_{tot}$ for the survey sensitivities is an approximation, since
$S_{lim}$ does not equal $5 \sigma_c$ in the general case, as discussed
in Sect.~\ref{confusionlimitsformips}.

The bottom part of Tab.~\ref{tab:total_noise_mips} gives the fluxes that will
limit the surveys. They are computed by using the approximation given
by the quadratic sum $\sqrt{5 \sigma_p^2 + S_{lim}^2}$, which provides
a smooth transition between the regime dominated by photon/detector
noise (24 $\mu m$) and the regime dominated by confusion noise (160 $\mu m$).
These values are taken to be the baseline for the further
discussions.
The final sensitivity for the 65 Sq. Deg. the SWIRE Legacy survey will
be the same as the GTO Shallow survey. 
The deep GTO surveys will be almost 4 times more sensitive (photon
noise) that the shallow ones, on about 2.5 Sq. Deg.; in the far
infrared, the confusion will nevertheless limit the final sensitivity. 
For the GOODS Legacy program, with 10h integration per sky pixel at 24
$\mu m$ on 0.04 Sq. Deg., we expect a final sensitivity of 54 $\mu$Jy
at 24 $\mu m$.

It is beyond the scope of this paper to investigate the properties of
the galaxy cluster targets of the SIRTF GTO program and to make
predictions, but in these fields, the confusion limits will
significantly be reduced due to the gravitational lensing by a
foreground rich cluster, which increases both the brightness and mean
separation of the background galaxies.
This effect has already been exploited successfully in the
mid-infrared (e.g. ISO, \cite{altieri99}) and in the submillimeter
(e.g., SCUBA Lens Survey, \cite{smail2002}). The SIRTF GTO program
will apply the same strategy in the mid- and far-IR. The lensed area
of the proposed GTO program is expected to cover 90 Sq. arcmin (E. Egami,
Private Communication).

Other effects, not included in this analysis, might slightly degrade
the final sensitivity of the maps, especially on Ge:Ga detectors at 70
and 160 $\mu m$; these effects, well characterized on the ground,
can probably be corrected with an accuracy of a few percent using data
redundancy and a carefully designed pipeline \cite[]{gordon2003}.
The effects are: stimulator flash latents (the amplitude is less than $
3\%$ and the exponential decay time constant is in the range 5-20 s),
transients, responsivity changes (tracked with the stimulator
flashes every 2 minutes), and cosmic ray hits related noise.
The final sensitivity will be measured in the first weeks of
operation, during the In Orbit Checkout and Science Verification
phases. 

%
\clearpage
\begin{deluxetable}{lccc}
\tablewidth{0pt}
\tablecaption{Number of Expected Sources in the MIPS Surveys, and
Fraction of the CIB that will be resolved into sources (assuming that all
sources are unresolved). Characteristics of the Surveys is given in
Tab.~\ref{tab:gto_mips}.\label{tab:number_sources_fraction_cib}}
\tablehead{
\colhead{} &\colhead{24 $\mu m$} & \colhead{70 $\mu m$} & \colhead{160 $\mu m$}
}
\startdata
\multicolumn{4}{l}{Number of Expected Sources}\\ \tableline
	Shallow    & $2.0 \times 10^4$   & $1.3 \times 10^4$    & $2.8\times 10^3$ \\
	Deep       & $2.5 \times 10^4$   & $5.8 \times 10^3$    & $1.4\times 10^3$ \\
	Ultra Deep & $3.7 \times 10^2$   & $49$     		& -- \\	
	SWIRE      & $1.5 \times 10^5$   & $1.0 \times 10^5$    & $2.2 \times 10^4$ \\
	GOODS      & $8.4 \times 10^2$   &  --   & -- \\ [+5pt]
\tableline
\multicolumn{4}{l}{Fraction of Resolved CIB}\\ \tableline
	Shallow    &  35\%  & 46\%   & 18\% \\
	Deep       &  58\%  & 54\%   & 23\% \\
	Ultra Deep &  68\%  & 54\%   & -- \\
	SWIRE      &  35\%  & 46\%   & 18\% \\
	GOODS      &  69\%  & --     & -- \\[+5pt]	
\enddata
\end{deluxetable}

%
\begin{figure}
\plotone{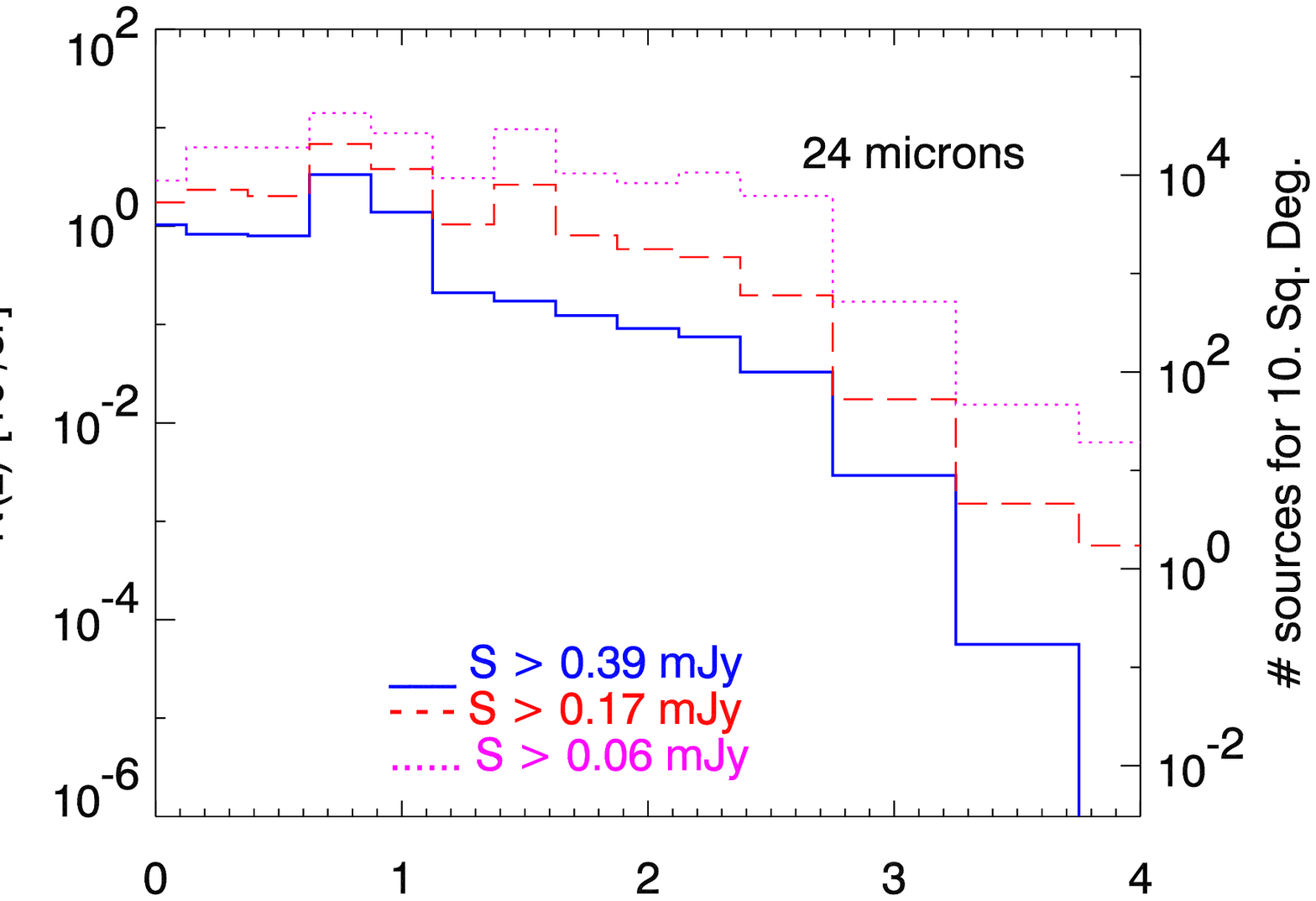}
\vspace{.1cm}
\caption{\label{fig:zdistrib_0024_hdole} Redshift
Distribution at 24 $\mu m$ with MIPS. Solid Line: Shallow Survey;
Dash: Deep Survey; Dot: Ultra Deep Survey. The flux limits
are listed in Tab.~\ref{tab:total_noise_mips}. Left axis gives the
source density (number of sources for the particular bin sizes shown),
right axis gives the number of sources in a 10. Sq. Deg. field.}
\end{figure}

%
\begin{figure}
\plotone{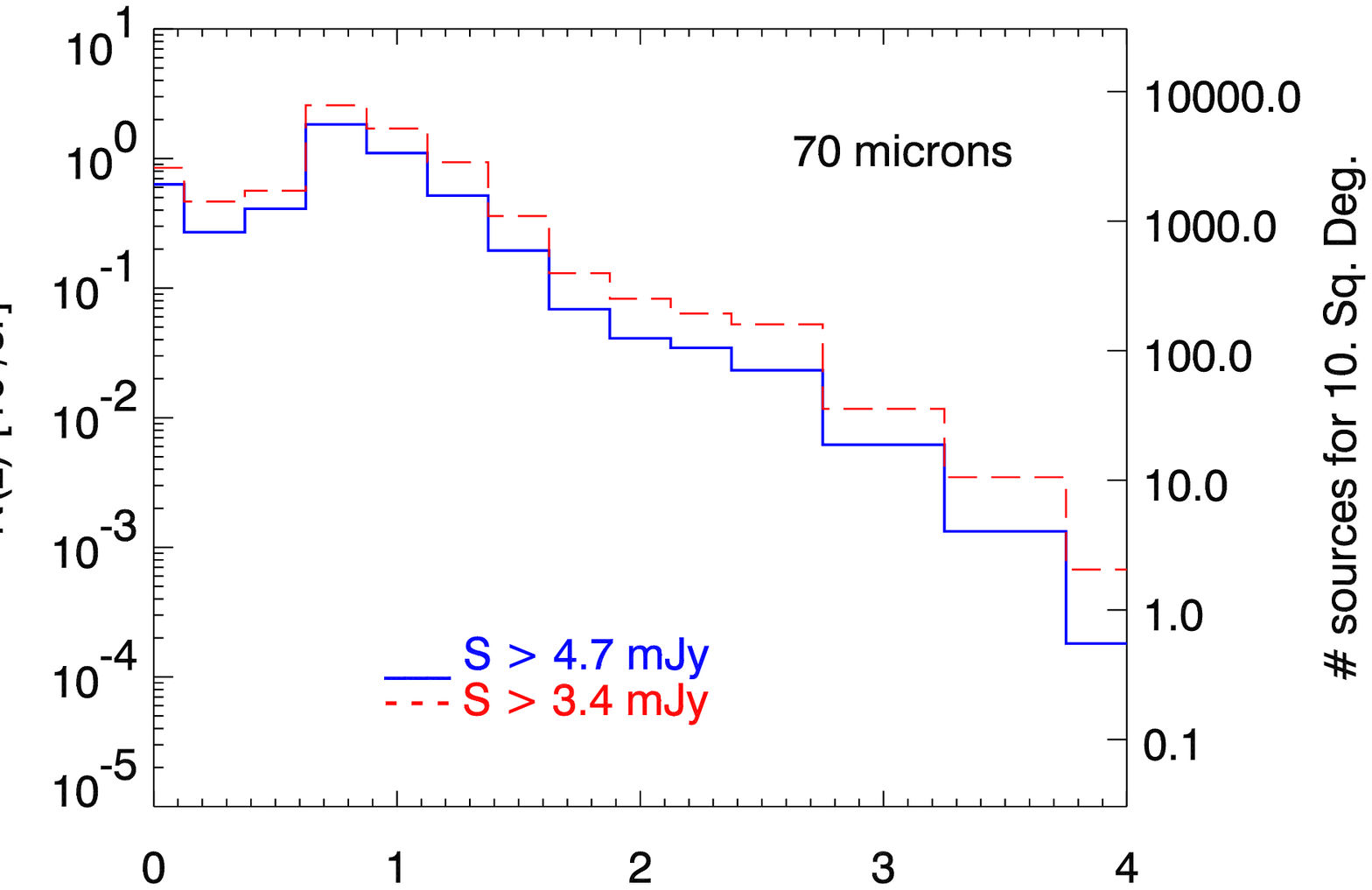}
\vspace{.1cm}
\caption{\label{fig:zdistrib_0070_hdole} Redshift
Distribution at 70 $\mu m$ with MIPS. Solid Line: Shallow Survey;
Dash: Deep Survey; Dot: Ultra Deep Survey. The flux limits
are listed in Tab.~\ref{tab:total_noise_mips}. Left axis gives the
source density (number of sources for the particular bin sizes shown),
right axis gives the number of sources in a 10. Sq. Deg. field.}
\end{figure}

%
\begin{figure}
\plotone{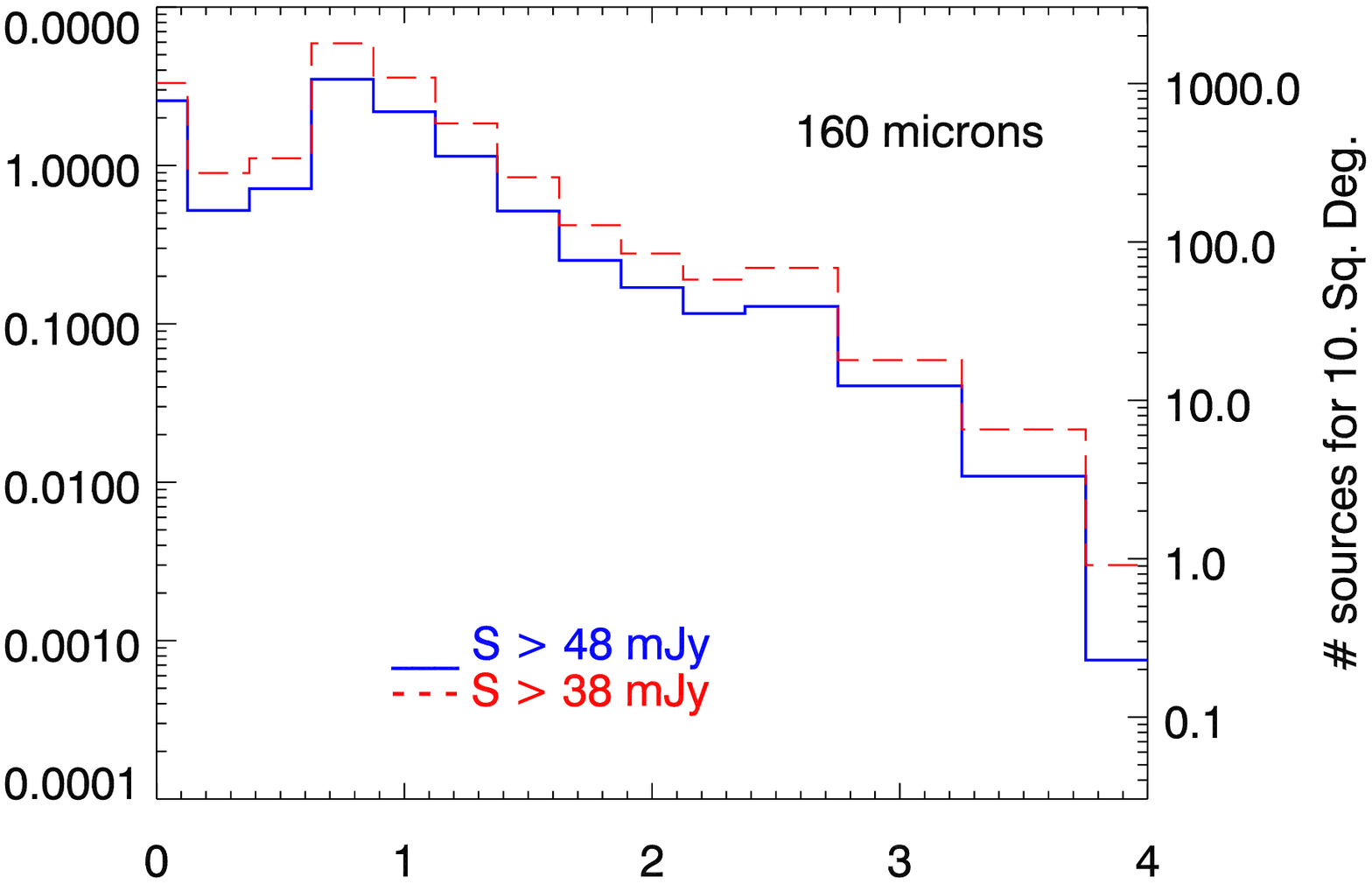}
\vspace{.1cm}
\caption{\label{fig:zdistrib_0160_hdole} Redshift
Distribution at 160 $\mu m$ with MIPS. Solid Line: Shallow Survey;
Dash: Deep Survey. The flux limits are listed in
Tab.~\ref{tab:total_noise_mips}. Left axis gives the source density
(number of sources for the particular bin sizes shown), 
right axis gives the number of sources in a 10. Sq. Deg. field.}
\end{figure}

%
\begin{figure}
\plotone{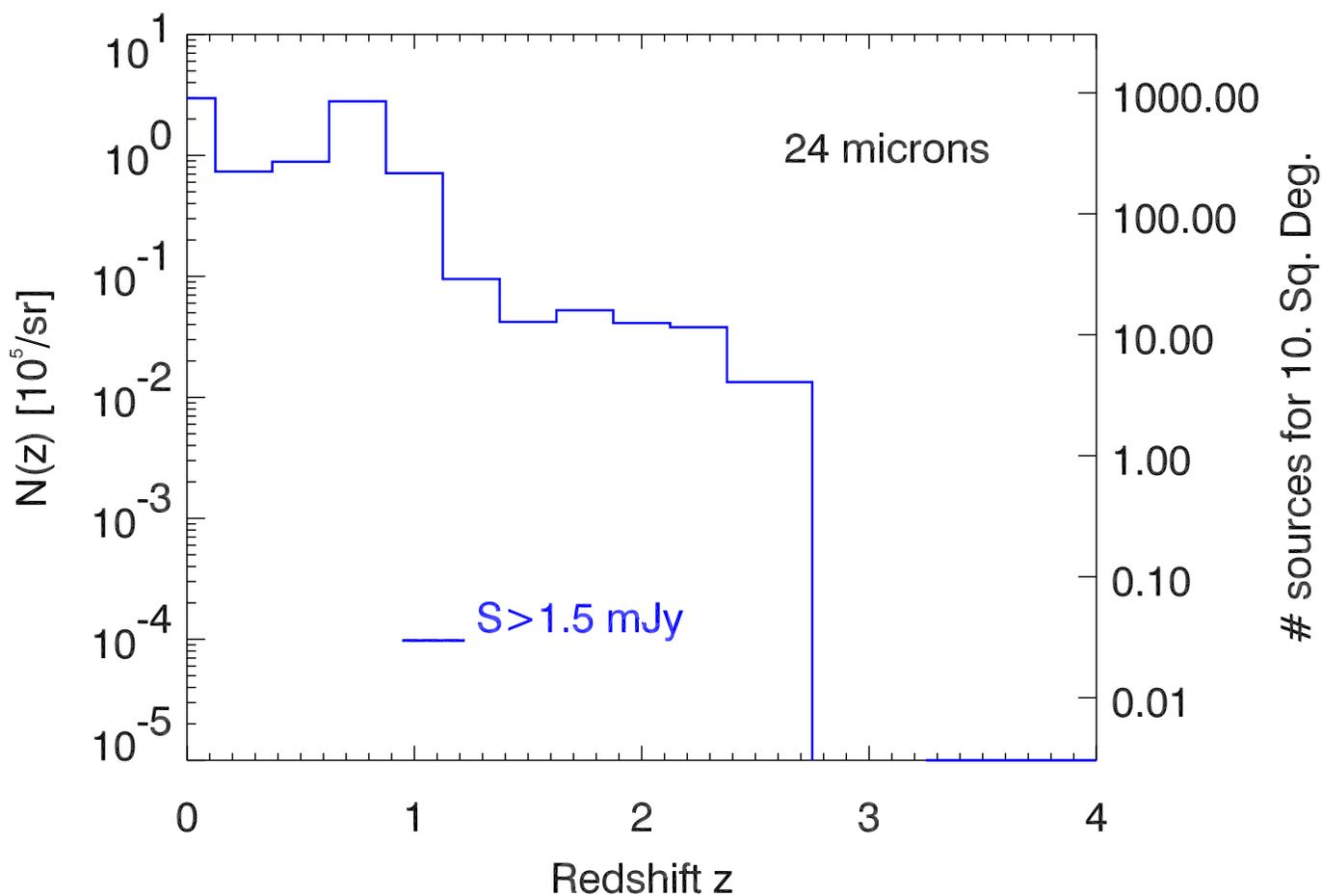}
\vspace{.1cm}
\caption{\label{fig:zdistrib_0024_hdole_IRS} Redshift
Distribution at 24 $\mu m$ with MIPS for sources
brighter than 1.5 mJy and 0.75 mJy, allowing a spectroscopic follow-up with
IRS. Left axis gives the source density (number of sources for the
particular bin sizes shown), right axis gives the number of sources in
a 10. Sq. Deg. field. In the proposed 9 Sq. Deg. shallow survey, we
would expect 2100 and 7200 sources to the 1.5 and 0.75 mJy depth respectively.} 
\end{figure}

%
\begin{figure}
\plotone{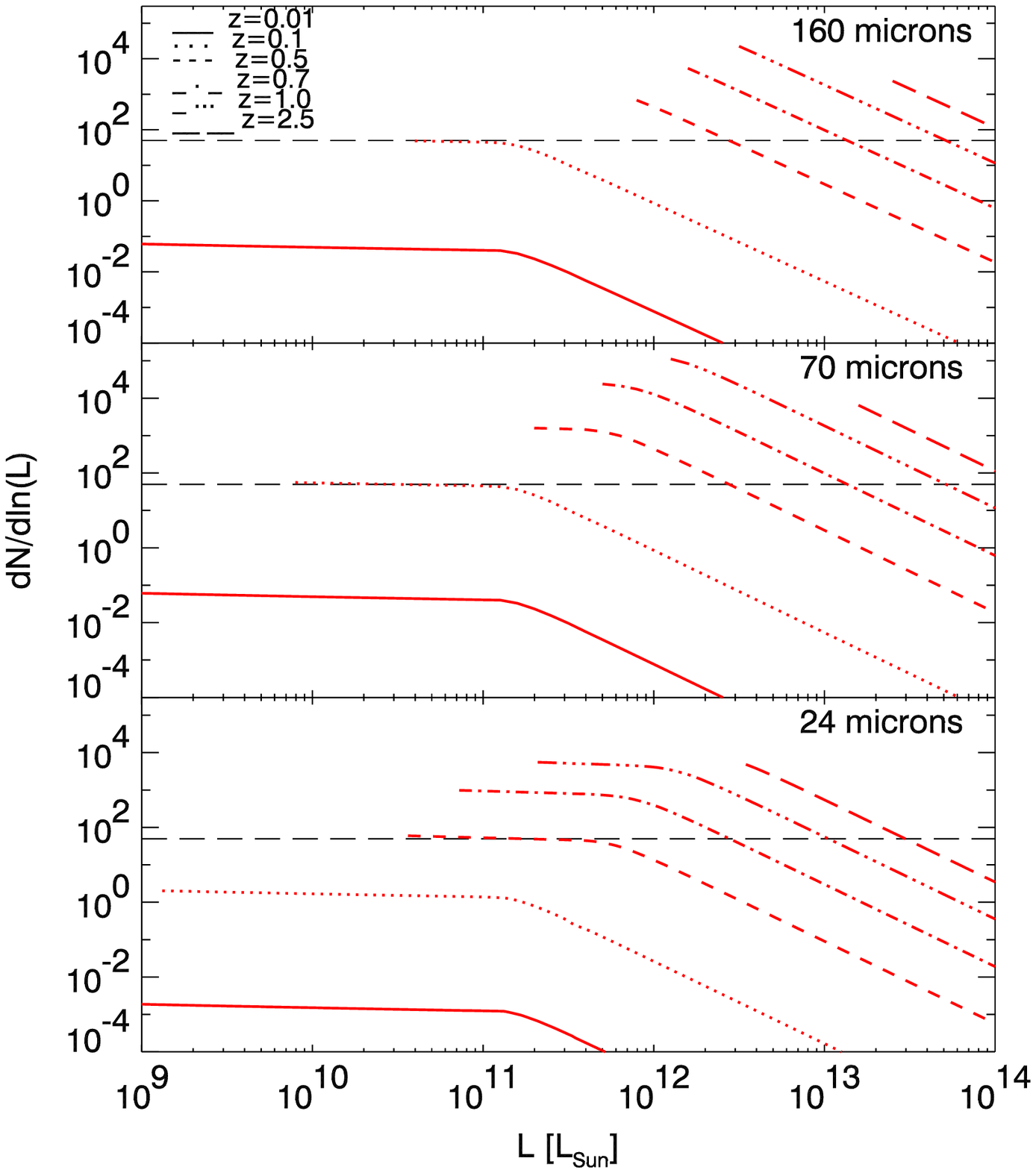}
\caption{\label{fig:dndlnl} Number of starburst galaxy per
logarithmic luminosity bin ($\Delta \ln{L} / \ln{L} = 0.1$)
that can be detected at different redshifts (with a $\Delta$z/z=0.5).
Top: 160 $\mu$m survey of 80 Sq. Deg. (surface covered by SWIRE and
the GTO) limited by the confusion at 48 mJy. Middle: 70 $\mu$m survey
of 80 Sq. Deg. limited by the confusion at 4.7 mJy. Bottom: 24 $\mu$m
survey of 2.46 Sq. Deg. down to 112 $\mu$Jy (GTO Deep Survey).
The horizontal dash line shows the 50 sources needed in a $\Delta$z/z=0.5
bin and $\Delta \ln{L} / \ln{L} = 0.1$ bin for reconstructing the
Luminosity Function (LF). 
From bottom left to upper right, redshift bins centered at
z=0.01, 0.1, 0.5, 0.7, 1.0, 2.5.}
\end{figure}

%
%
\section{Resolved Sources: Redshift Distributions, Luminosity
Function, Resolution of the CIB}
\label{sect:resolvedsources}
%
\subsection{Source Density and Redshift Distributions}
Many resolved sources are anticipated in the MIPS surveys: for
instance, we expect at 160 $\mu m$ a number of sources more than an
order of magnitude higher than those detected by ISO, due to both a fainter
detection limit and a larger sky coverage. Tab.~\ref{tab:number_sources_fraction_cib}
gives the number of sources for the GTO and Legacy surveys.

The redshift distributions of the surveys are plotted in
Fig.~\ref{fig:zdistrib_0024_hdole} for 24 $\mu m$, 
Fig.~\ref{fig:zdistrib_0070_hdole} for 70 $\mu m$, and
Fig.~\ref{fig:zdistrib_0160_hdole} for 160 $\mu m$.
At 24 $\mu m$, the deepest fields will allow us to probe the dust
emission of sources up to redshift of 2.7. At higher redshifts, the
7.7 $\mu m$ PAH feature causes a fall in the K-correction and thus a
decrease in the observed flux close to the sensitivity limit. This is
similar to the drop observed with ISOCAM at 15 $\mu m$ for sources
lying at redshift 1.4. (This does not exclude to detect the stellar
emission at larger redshifts; this is outside the scope of this paper).

At 70 $\mu m$, the redshift distribution peaks at 0.7, with a tail
extending up to redshift 2.5.
At 160 $\mu m$, the redshift distribution is similar to that at 70 $\mu m$.
In the far infrared, MIPS surveys will probe extensively the largely
unexplored $1 < z < 2.5$ regime.

%
\subsection{Spectra with IRS}
Spectra of some high redshift sources will be taken with IRS on board
SIRTF (as part of the IRS GTO program). With a sensitivity limit of
1.5 mJy at 24 $\mu m$ and maybe 0.75 mJy (Weedman, Private
Communication), a few dozen sources at redshift greater than 2 will be
observed. Fig.~\ref{fig:zdistrib_0024_hdole_IRS} shows the predicted
redshift distribution at 24 $\mu m$ for the Shallow Survey
of the sources that might be followed-up in 
spectroscopy by IRS. 

%
\subsection{Luminosity Function Evolution}
In addition to the photometric redshifts of a large number of sources,
and spectroscopic redshift following identifications,
building the luminosity function of the sources as a function of redshift
will be one of the key results of the SIRTF surveys.
We show in Fig.~\ref{fig:dndlnl} the source density per logarithmic
luminosity bin and per redshift bin expected in
the MIPS surveys.

The source density of starburst galaxies is given per
logarithmic luminosity bin 
(of $\Delta \ln{L} / \ln{L} = 0.1$) and for
redshift bins (of width $\Delta$z/z=0.5) ranging from z=0.01 to z=2.5.
The survey sensitivity cuts the distributions at low luminosities.
The size of the surveys limits the ability to derive the LF at high
luminosities. The limit of 50 sources per z and L bins is also
shown. This limit ensures an statistical accuracy of 14\% on the
LF for each luminosity and redshift bin; averaging over five bins in
luminosity (thus getting $\Delta \ln{L} / \ln{L} = 0.5$), allows to
reach an accuracy of 6\%. 

At 160 $\mu m$ (top of Fig.~\ref{fig:dndlnl}), with a 48 mJy limiting
flux and a coverage of 80 Sq. Deg., corresponding to the surface
covered by all the legacy and GTO extragalactic programs, the MIPS
data should allow to reconstruct the LF 
of some ULIRGs ($10^{12}L_{\odot} < L < 3 \times 10^{12}L_{\odot}$)
in the $0.5 < z < 0.7$ range,
of the $3 \times 10^{12}L_{\odot} < L <  10^{13}L_{\odot}$ galaxies 
in the $0.5 < z \lesssim 1$ range, 
and of the HyLIGs \cite[]{morel2001} ($L > 3 \times 10^{13}L_{\odot}$) in the
$1 \lesssim z \lesssim 2.5$.

At 70 $\mu m$ (middle of Fig.~\ref{fig:dndlnl}), with a 4.7 mJy limiting
flux and a coverage of 80 Sq. Deg., the sensitivity in the wide
and shallow surveys allows to probe in addition the
$3 \times 10^{11} < L < 10^{12}L_{\odot}$ 
sources at z=0.5, and the full range $10^{12} < L < 10^{13}L_{\odot}$
for sources at $0.7 < z < 1$.

At 24 $\mu m$, the situation is very similar to that at 70 $\mu m$ for
these shallow surveys (limiting flux of 390 $\mu$Jy), except a slightly better sensitivity to
galaxies with $L \sim 10^{11}L_{\odot}$ around z=0.5.
Concerning deeper and narrower surveys at 24 $\mu m$ (limiting flux of
112 $\mu$Jy), like the GTO deep surveys, the sensitivity to lower
luminosities galaxies at higher redshifts is better
(bottom of Fig.~\ref{fig:dndlnl}). In the 
redshift range 0.5 to 2.5, the gain in sensitivity compared to 70 $\mu
m$ allow to probe lower luminosities galaxies, by a factor of $\sim 5$.

%
%
\subsection{Resolving the CIB}
\label{resolvingcib}
To compute the fraction of the CIB that will be resolved into sources,
one has to consider the apparent size of the
galaxies. \cite{rowan-robinson74} give the formalism to deal with
resolved and extended sources. To simplify the problem, one might
check if all the sources are point sources. For MIPS, the underlying
assumption about the physical size of the objects is that it is smaller
than 40 kpc, corresponding to less than the FWHM at 24 $\mu m$ at
$z>1$.
Indeed, most of the galaxies observed in the HDF-N with
NICMOS exhibit structures smaller than 25 kpc ($\simeq 3$ arcsec) in
the redshift range  $z = 1$ to $2$ \cite[]{papovich2002b}. The objects
are thus smaller than the MIPS beam sizes and won't be resolved. This
might not be the case for IRAC. The closer resolved objects give a
negligible contribution to the background anyway. 

The fraction of the CIB resolved into discrete sources is
given in Tab.~\ref{tab:number_sources_fraction_cib}. 
MIPS will resolve at most
69\%, 54\% and 24\% of the CIB at  24, 70 and 160
$\mu m$ respectively. This is an improvement by a factor of at least 3 of the CIB
resolution in the FIR over previous surveys (e.g. at 170 $\mu m$, \cite{dole2001}).
At 24 $\mu m$, most of
the CIB will be resolved, as ISOCAM did at 15 $\mu m$
\cite[]{elbaz2002a}, but with a much wider and deeper redshift coverage.

%
\subsection{Conclusion: Multiwavelength Infrared Surveys}
In the far infrared range, the most promising surveys appear to be the large
and shallow ones, because 1) the large number of detected sources is a key
to have a statistically significant sample, and 2) the confusion level and
the sensitivity is enough to probe sources in the redshift range
from 0.7 to 2.5. Together with a significant resolution of the CIB at
70 and 160 $\mu m$ (46 and 18\% respectively), the surveys will
tremendously improve our knowledge of the sources that ISO could not
detect. 
In the mid infrared range, where the confusion is negligible, the need
for deeper surveys is striking. The Deep and Ultra Deep surveys will
resolve most of the CIB at 24 $\mu m$, allowing not only to study
populations from z=0 to z=1.4 (like ISO did), but also the population that lie at z
between 1.5 to 2.7, with unprecedented accuracy \cite[]{papovich2002}.
All these multi-wavelength surveys (GTO \& Legacy programs)
will thus probe for the first time a
population of infrared galaxies at higher redshift, allowing to
characterize the evolution, derive the luminosity function evolution,
constrain the nature of the sources, as well as deriving the unbiased
global star formation rate up to z $\sim$ 2.5. 

%
\begin{figure}
\plotone{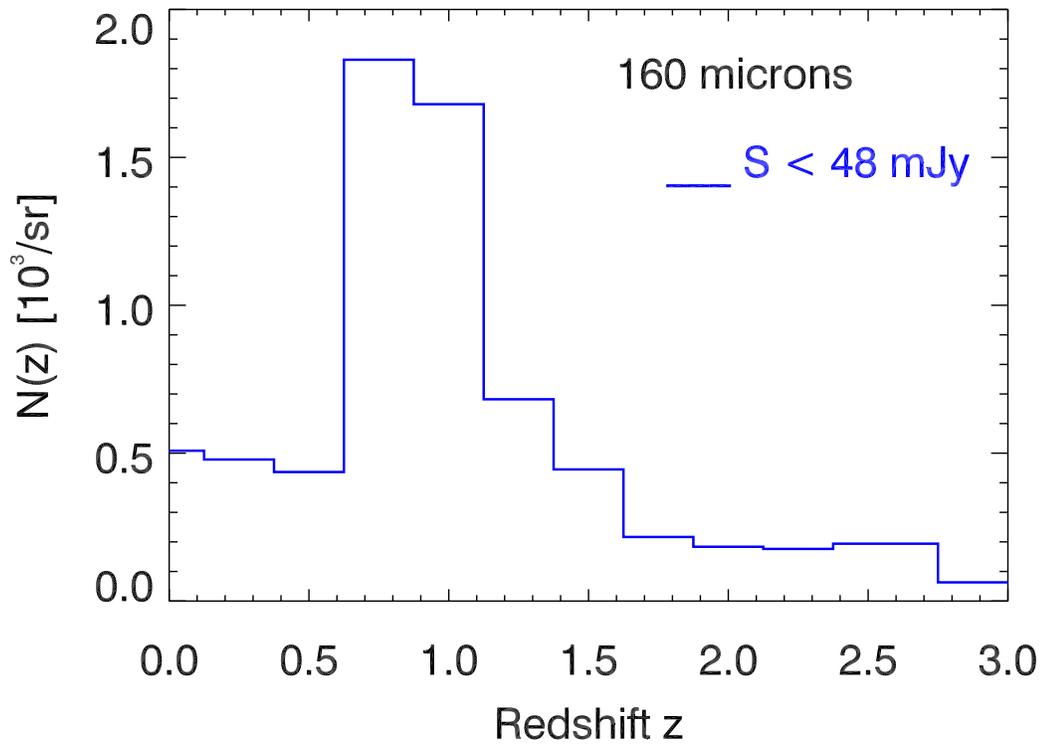}
\caption{\label{fig:zdistrib_0160_hdole_fluc} Redshift
Distribution of the sources below 48 mJy creating the fluctuations,
at 160 $\mu m$ with MIPS. The number of sources is shown for the
particular bin sizes.} 
\end{figure}

%
\begin{figure}
\plotone{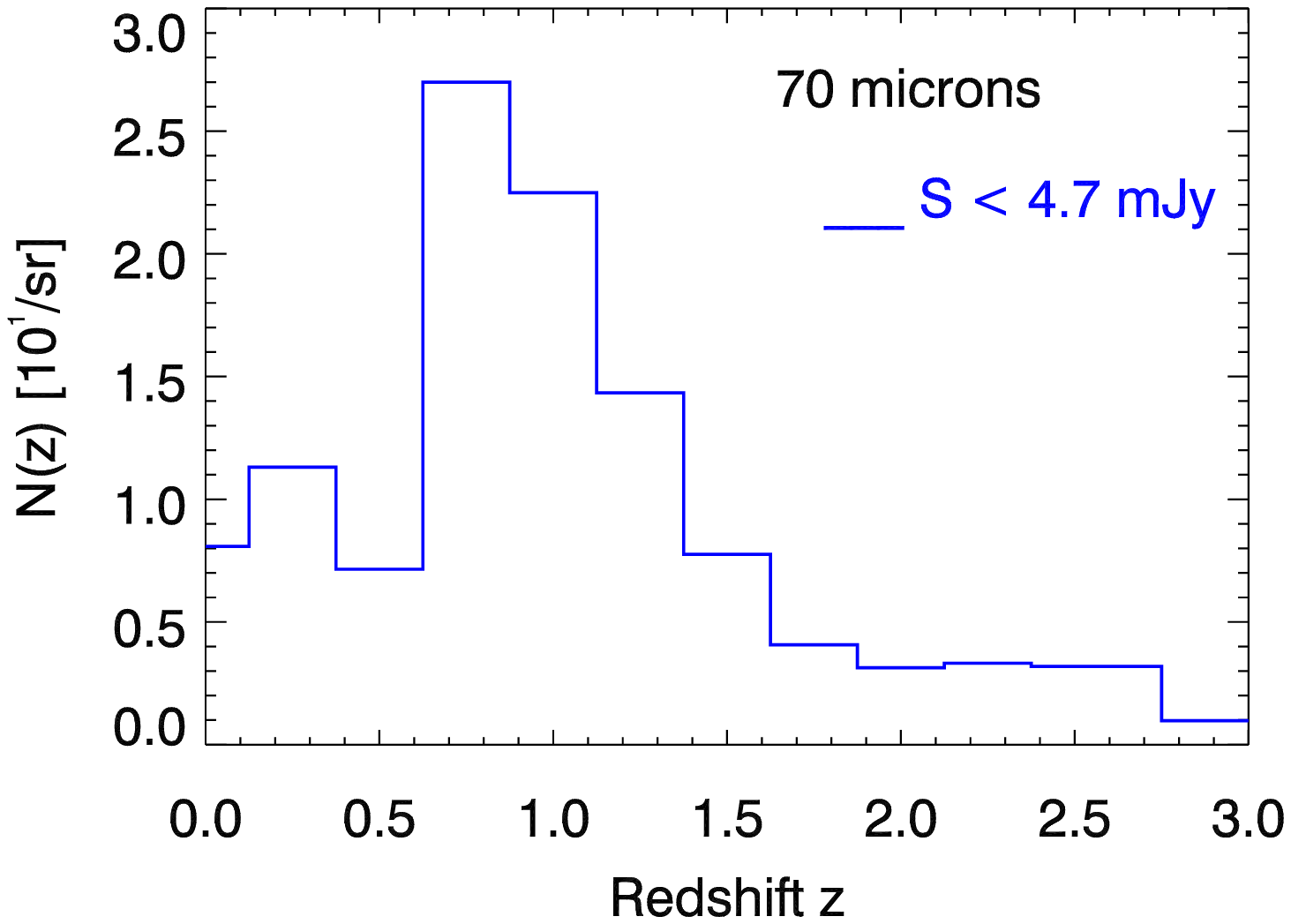}
\caption{\label{fig:zdistrib_0070_hdole_fluc} Redshift
Distribution of the sources below 4.7 mJy creating the fluctuations,
at 70 $\mu m$ with MIPS. The number of sources is shown for the
particular bin sizes.} 
\end{figure}
%

%
%
\section{Unresolved Sources: Fluctuations of the Cosmic Infrared Background}
\label{sect:unresolvedsources}
%
\subsection{Fluctuation Level and Redshift Distributions}
\label{sect:fluctuations}
Sources below the detection limit of a survey create fluctuations. If
the detection limit does not allow to resolve the sources
dominating the CIB intensity, characterizing these fluctuations
gives very interesting informations on the spatial correlations of
these unresolved sources of cosmological significance.
The far infrared range is ``favored''
for measuring the fluctuations, because data are available with very
high signal to detector noise ratios, but limited by the confusion;
on the other hand, the confusion limits the possibility to detect faint
resolved sources and leaves the information about faint sources hidden in
the fluctuations. 
The study of the CIB fluctuations is a rapidly evolving
field. After the pioneering work of \cite{herbstmeier98} with ISOPHOT,
\cite{lagache2000a} discovered them at 170 $\mu m$ in the 
FIRBACK data, followed by other works at 170 and 90 $\mu m$
\cite[]{matsuhara2000,kiss2001,puget2000}, and at 60
and 100 $\mu m$ by \cite{miville-deschenes2002} in the IRAS data.

Our model reproduces the measured fluctuation levels within a factor
1.5 between 60 and 170 $\mu m$ \cite[]{lagache2002}.
For MIPS, we predict that the level of the fluctuations is
6930 Jy$^2/sr$ at 160 $\mu m$ for $S_{160} < 48$ mJy, and
is 113 Jy$^2/sr$ at 70 $\mu m$ for $S_{70} < 4.7$ mJy.

Our model gives access to the redshift distribution of the sources
dominating the observable fluctuations of the unresolved background. 
At 170 $\mu m$ (Fig.~12 from \cite{lagache2002}), the redshift
distribution of the contributions to the fluctuations peaks at
z=0.8, with a tail up to z $\sim$ 2.5, and there is a non negligible
contribution from local sources. This peak of this distribution is
similar to the one of the
$15 \mu m$ ISOCAM redshift distribution of resolved sources
\cite[]{elbaz2002a}, which are understood to represent a significant
fraction of the CIB. These sources observed at two different
wavelengths should tell us the same story about galaxy evolution. The
key point of studying the fluctuations in the far infrared is the
availability of large area surveys to exhibit the source clustering
properties; this is not yet possible with mid infrared data that need
to be taken with deeper (and thus less area coverage) exposures to probe
the same sources. Furthermore, a non negligible contribution comes
from higher redshifts. Extracting this component will be a challenge
requiring the use of all SIRTF bands. 

At 160 $\mu m$ (Fig.~\ref{fig:zdistrib_0160_hdole_fluc}), for the same
reasons, the distributions of the sources dominating the fluctuations
peaks at z=0.8, with a broad peak from z=0.7 to z=1.1. The tail
extends up to z $\sim$ 2.5, and the contribution of local sources is
less prominent than at 170 $\mu m$ with the ISOPHOT sensitivity.
At 70 $\mu m$ (Fig.~\ref{fig:zdistrib_0070_hdole_fluc}), the
distribution is similar to that at 160 $\mu m$, but with a factor of
three less source density, since the background is half resolved into
sources.

%
\subsection{Power Spectrum Analysis: Fluctuations and Source Clustering}
\label{sect:powerspectrum}
The Poisson component of the fluctuations of the CIB has been detected
in the FIR by \cite{lagache2000a} in the FIRBACK data, at spatial
frequencies (or wavenumbers) $0.25 < k < 0.6$ arcmin$^{-1}$. 
A preliminary study on larger fields seems to show that the source
clustering is present in the data as well \cite[]{puget2000}, and are
currently under investigation \cite[]{sorel2003}. However, to
accurately constrain the source clustering, larger fields than FIRBACK
are needed.
Since SIRTF will cover larger sky areas, the clustering should be
detected and measured in a power spectrum analysis similar to the one
done by \cite{lagache2000a} and \cite{miville-deschenes2002}.

We make an estimation of the spatial frequency range where the CIB
fluctuations will be detected in the large and shallow surveys at 160
$\mu$m, using our model; it does not include source clustering, we just
assume a Poisson distribution of the sources. The detectability of the
source clustering is addressed below. 

We use the same technique as \cite{lagache2000a} and
\cite{puget2000}; following their formalism, the power spectrum
measured on the map $P_{map}$, in the space of the detector, can be
written as follows:
\begin{equation}
\label{eq:pmapdet}
P_{map} = P_{noise} + ( P_{cirrus} + P_{sources}) \times W_k
\end{equation}
where $P_{noise}$, $P_{cirrus}$, $P_{sources}$ are the power spectra
of the photon/detector noise, the foreground cirrus, and the
extragalactic sources we are interested in respectively, and $W_k$ is
the power spectrum of the PSF. In this analysis, we want to exhibit
$P_{sources}$ and, for convenience,
$P_{cirrus}$.

\begin{figure}
\plotone{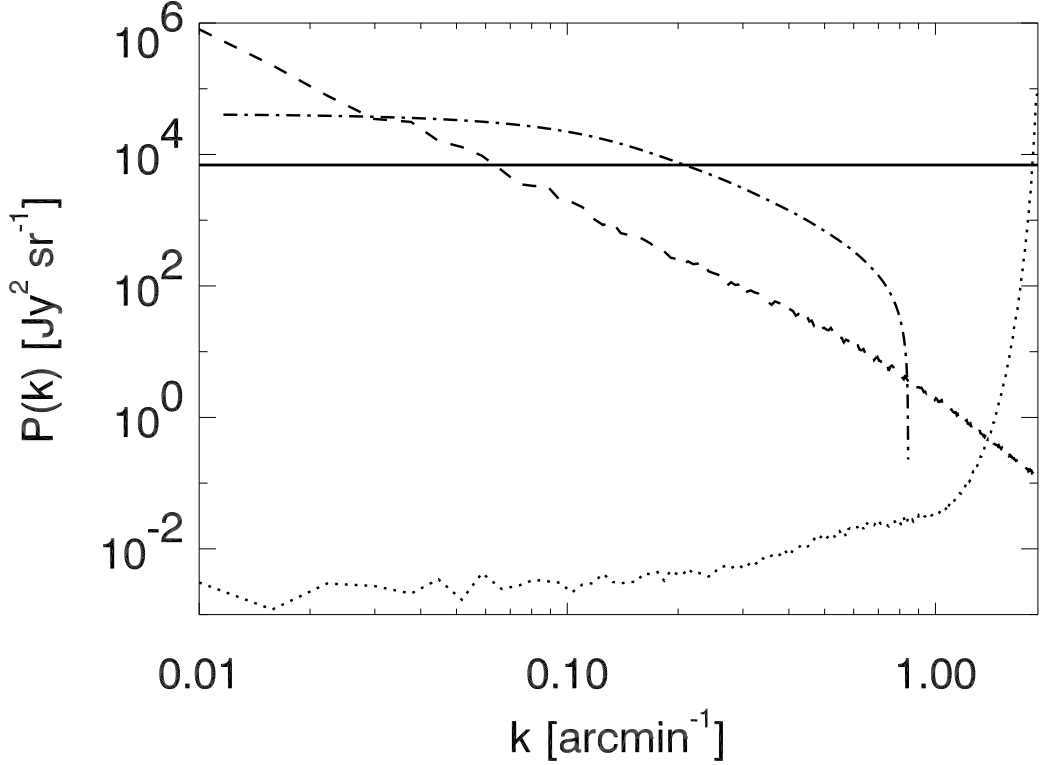}
\caption{\label{fig:predict_powspec} Theoretical power spectrum of a 5
Sq. Deg. field at 160 $\mu m$, illustrating the spatial
frequency range where the CIB fluctuations will be detected 
(see text Sect.~\ref{sect:powerspectrum}).
Solid line: level of CIB Poisson fluctuations created by sources
below 48 mJy predicted by our model: 6930 Jy$^2$/sr.
Dash: Foreground Cirrus, $P_{cirrus}$, with the $k^{-3}$ behaviour and
normalized at $10^6$ Jy$^2$/sr at $k=0.01$ arcmin$^{-1}$, representing
a column density of $N_{HI} = 10^{20}$ cm$^{-2}$.
Dot: White noise ($1 \sigma$ of 7 mJy) divided by the PSF, $P_{noise} / W_k $.
Dash-Dot: model source clustering of starburst galaxies
of \cite{perrotta2001} in the case of 170 $\mu m$. 
The wavenumber range of cosmological interest is thus from 0.07
to 1.3 arcmin$^{-1}$, where the CIB Poisson fluctuations are expected to be
detected; assuming the source clustering has the form predicted by
\cite{perrotta2001}, it will be detected in the wavenumber range from
0.04 to 0.2 arcmin$^{-1}$.}
\end{figure}

Fig.~\ref{fig:predict_powspec} shows a prediction for
the various components present at 160 $\mu m$ in a survey like the GTO
Shallow or SWIRE.
$P_{sources}$, the Poisson component for the fluctuations due to extragalactic
sources fainter than 48 mJy is shown as an horizontal line, at the
value of 6930 Jy$^2/sr$ predicted by the model (see
Sect.\ref{sect:fluctuations}). 
$P_{cirrus}$ is shown as a dash line, and follows a $k^{-3}$ power law
\cite[]{gautier92,miville-deschenes2002}. The normalization at $10^6$
Jy$^2/sr$ at $k=10^{-2}$ arcmin$^{-1}$ is typical of the faint cirrus
present in the cosmological fields of column density 
$N_{HI} = 10^{20}$ cm$^{-2}$. Finally, $P_{noise} / W_k$ is
plotted as a dot line. The noise is a white noise of $1\sigma$ of 7
mJy, typical for shallow surveys at 160 $\mu m$
(Tab.~\ref{tab:total_noise_mips}). 

To have an estimation of the spatial frequency range where the
Poisson fluctuations from the extragalactic component will be detected, one
has to consider the two limiting components: galactic cirrus at low
spatial frequencies, and photon noise plus PSF shape at large spatial
frequencies. 
It appears that the CIB Poisson fluctuations, or the fluctuations
created by faint extragalactic sources only, should be well detected
in the wavenumber range $0.07 < k < 1.3$ arcmin$^{-1}$. 

Taking into account the source clustering, we assume that it is
dominated by starburst galaxies with the form predicted by
\cite{perrotta2001}\footnote{Other predictions exist in the submm
range, but not specifically for 160 $\mu
m$ \cite[]{haiman2000,knox2001}. The source clustering is there
expected at the scales between 0.1 and 3$^{\circ}$. }.
This clustered component is plotted in
Fig.~\ref{fig:predict_powspec} as a dot-dash line, and has been
computed for 170 $\mu m$. This component
should be detected in the wavenumber range from
0.04 to 0.2 arcmin$^{-1}$. The cirrus limits the detection at
smaller wavenumbers, and is the main limitation for the source
clustering detection. The Poissonian component of extragalactic
sources limits the detection at larger wavenumbers. 

The large shallow surveys in the FIR are thus the most promising for
studying the fluctuations and estimating the source clustering
($0.04 < k < 0.2$ arcmin$^{-1}$).

%
%

%
\section{Conclusion}
\label{sect:conclusion}
In this work, we review the sources of noise expected in the
cosmological surveys to be conducted by MIPS: photon/detector
noise, cirrus noise, and confusion noise due to extragalactic sources.
Using the Lagache, Dole, \& Puget (2002) model, as well as the latest
knowledge of the MIPS pre-flight characteristics (in particular the
photon/detector noise properties and the beam shapes), we predict
the confusion levels, after a detailed discussion on the criteria. In
particular, we show that in general the criterion depends on the shape of the
source counts and the solid angle of the beam (directly related to the
telescope and detector pixel size).
SIRTF is about to probe a new regime in the source counts, where a
significant fraction of the CIB is resolved and the counts begin to
flatten. We thus discuss the classical rules of determining the
confusion level (essentially valid for IRAS or ISO), and we show that it is
wise to compare the photometric and source density criteria for
predicting the confusion level. We found $S_{lim}$ to be 50 $\mu$Jy,
3.2 mJy and 36 mJy at 24, 70 and 160 $\mu m$ respectively, consistent
with ISO data or other works. 

We compute the final sensitivity of the MIPS surveys, the GTO (guaranteed
time) and two Legacy programs (SWIRE and GOODS), predict the number
of sources, and give the redshift distributions of the detected sources at 24,
70 and 160 $\mu m$. The deepest surveys should detect the dust
emission of sources up to z=2.7 at 24 $\mu m$ (the redshifted 7.7 PAH
feature causes a drop of detectability at higher redshifts), and up to
z=2.5 at 70 and 160 $\mu m$.  
This corresponds to a resolution of the CIB into discrete sources of
69, 54 and 24\% at 24, 70 and 160 $\mu m$ respectively.
We estimate that in the shallow surveys, the sources will be detected in a
sufficient number in redshift bins to reconstruct the luminosity
function and its evolution with redshift with a 14\% (or better) accuracy, as
follows:
most of the $L > 10^{12}L_{\odot}$  in the
$0.5 \lesssim z \lesssim 1$ in the FIR range and
most of the $L > 10^{11}L_{\odot}$  in the
$0.5 \lesssim z \lesssim 1$ in the MIR range,
and all the $L \gtrsim \times 10^{13}L_{\odot}$ sources for
$z \simeq 2.5$ in the MIR and FIR range. We also show that at 24 $\mu
m$, deeper and narrower surveys will considerably increase the
sensitivity to lower luminosity galaxies.

We also explore some characteristics of the unresolved sources at
long wavelength, among which the redshift distribution of the
contribution to the background fluctuations at 70 and 160 $\mu m$. It peaks
at $z \sim 0.8$, consistent with our present understanding of the
main contribution to the CIB. We estimate the wavenumber range
where the large FIR surveys will be able to measure the fluctuations
of the Poisson component in a power spectrum analysis is
$0.07 < k < 1.3$ arcmin$^{-1}$. With some assumption about the source
clustering, we show that it could be detected in the wavenumber range
$0.04 < k < 0.2$ arcmin$^{-1}$.

We emphasize the complementary role of large and shallow surveys in the
far infrared and smaller but deeper surveys in the mid infrared. The
MIR surveys allow to probe directly faint sources, and FIR surveys
allow to access the statistical properties of the faint population,
mainly through CIB fluctuation analysis. With the various sky area
coverage and depth, the MIPS surveys (together with IRAC data helping
to estimate the photometric redshifts) will 
greatly improve our understanding of galaxy evolution by providing
data with unprecedented accuracy in the mid and far infrared range.

%
\acknowledgments
We thank George \& Marcia Rieke for many interesting discussions, as
well as Almudena Alonso-Herrero, Eiichi Egami \& Casey Papovich.
We also thanks Petri Vaisanen for helpful discussions about the 8
$\mu$m source counts, and for having provided us with their
predictions, as well as Francesca Perrotta and Manuela 
Magliocchetti for having provided us with an electronic version of
their model. We also appreciated remarks from Alberto Franceschini and
Kevin Xu, and discussions with Dario Fadda and Dan Weedman. We thank
the referee for the constructive comments and suggestions. 
HD thanks the funding from the MIPS project, which is supported by
NASA through the Jet Propulsion Laboratory, subcontract \# P435236,
and the Programme National de Cosmologie and the Centre National
d'Etudes Spatiales (CNES) for travel funding.\\

%

\end{document}